\documentclass[reprint,nofootinbib,superscriptaddress,amsmath,amssymb,aps,prb,floatfix]{revtex4-2}
\usepackage[T1]{fontenc}
\usepackage{lmodern}
\usepackage{graphicx}
\usepackage{xcolor}
\usepackage{url}
\usepackage[colorlinks=true,linkcolor=blue!50!black,citecolor=blue!50!black,urlcolor=blue!50!black]{hyperref}
\newcommand{\sdos}{\textsc{sdos}}
\newcommand{\Tr}{\mathrm{Tr}}
\newcommand{\ie}{\emph{i.e.}}

\begin{document}

\title{\sdos: spectral densities of large sparse Hermitian and generalized eigenproblems}

\author{David A.\ Drabold}
\affiliation{Department of Physics and Astronomy, Ohio University, Athens, Ohio 45701, USA}

\date{November 8, 2026}

\begin{abstract}
\sdos\ is an open Python package that computes the density of states (DOS) and related spectral quantities of very large sparse Hermitian matrices $A$, and of generalized problems $A\psi=\lambda B\psi$, from Chebyshev moments estimated with random vectors. The density is reconstructed by the kernel polynomial method (KPM, the default) or by maximum entropy (Maxent); the same interface gives band energies, Fermi levels, local and projected densities, harmonic thermodynamics, exact HOMO--LUMO gaps and eigenpairs in a spectral window. Interfaces cover electronic (empirical tight-binding, DFTB, SIESTA) and vibrational (matrix-free LAMMPS Hessians) problems.
I checked most ingredients against dense diagonalization or an independent code; the exceptions are listed in Sec.~\ref{sec:scaling}.
On a workstation (4 threads, 19~GB) the DOS of an 884\,736-atom amorphous silicon model (a $6^3$ tiling of a 4096-atom cell; 3.5 million orbitals; the orthogonal empirical tight-binding model of Kwon \emph{et al.}) takes 71~s ($M=100$, $R=2$, no exact moments; 383~s with the default exact moments), and a matrix-free 2\,097\,152-atom crystal 10~min in 1.8~GB. A $17^3$ replica of a 216-atom amorphous cell (1\,061\,208 atoms) agrees with exact Bloch spectra: electronic DOS in 229~s, matrix-free ACE vibrational DOS in 2.2~h on one GPU.
Shift-invert, incomplete-Cholesky setup, exact moments of high order and Maxent at high moment counts do not scale. The underlying methods are due to Silver, R\"oder, Kress, Voter, Drabold, Sankey and collaborators; the Introduction states what is added.
\end{abstract}

\maketitle

\section{Introduction}
\label{sec:intro}

The method of this report applies to \emph{any} large sparse Hermitian matrix $H$ (and to any matrix-free linear operator, and to the generalized problem $A\psi=\lambda B\psi$ with $B$ positive definite). It computes the spectral density $\rho(E)=N^{-1}\Tr\,\delta(E-H)$, where $E$ is the energy (the spectral variable) and $N$ the dimension, and quantities derived from it. Nothing in the algorithm depends on where the matrix comes from. I illustrate the method with electronic and vibrational problems, because I know them, because exact references exist for them, and because the density of states determines band energies, Fermi levels and thermodynamics and, through its local variants, much of the interpretation of electronic and vibrational structure in disordered and nanostructured materials. These applications are examples of a more general tool.
For matrices of dimension $N\sim10^5$--$10^7$, as arise for amorphous networks, nanostructures and large supercells, diagonalization ($O(N^3)$ time, $O(N^2)$ memory) is out of reach. The \emph{moments} $\mu_n=N^{-1}\Tr\,T_n(\tilde H)$ of $\rho$ with respect to Chebyshev polynomials of the scaled operator, on the other hand, cost $O(N)$ operations each. They come from matrix--vector products with a few random vectors, because the trace can be replaced by a stochastic estimator \cite{skilling,hutchinson}.
Two routes lead from moments to a density. The kernel polynomial method (KPM) \cite{silver1994,silver1996,kpm} is a damped Chebyshev series; it is linear, easy to use, and positive with the Jackson kernel \cite{jackson}. The principle of maximum entropy \cite{mead,turek,drabold1993} finds the positive density of maximal entropy that reproduces the moments. It has an information-theoretic foundation \cite{jaynes1978} and, for the same moments, resolves sharp features better.
Both belong to the family of linear-scaling electronic-structure methods reviewed by Goedecker \cite{goedecker} and Ordej\'on \cite{ordejon1998}; related moment methods are the recursion method \cite{haydock}, the plane-wave moments method \cite{wang1994} and Fermi-operator expansions \cite{goedeckercolombo}, and modern surveys of stochastic spectral-density estimation are in Ref.~\cite{lin}.

Electronic problems in a non-orthogonal basis and vibrational problems with unequal masses are \emph{generalized} eigenproblems, $A x=\lambda Bx$.
R\"oder, Silver and Kress \cite{roder1996} showed that KPM extends to this case by applying $B^{-1}A$ with a preconditioned conjugate-gradient (PCG) solve of $B$ for every Chebyshev step, and demonstrated it for a titanate/platinum interface and amorphous diamond; R\"oder, Silver, Drabold and Dong \cite{roder1997} treated amorphous diamond in the same way.
\sdos\ implements this idea and adds alternatives for applying $B^{-1}$ (block-Jacobi, incomplete Cholesky, algebraic multigrid, and a fixed Chebyshev polynomial $p$ with $p(B)\approx B^{-1}$, which replaces the inverse by a matrix polynomial so that the recursion is applied to one fixed, explicit operator $p(B)A$ and needs no inner solves), a Maxent reconstruction, exact low-order moments, spectral windows, projections and a set of interfaces to tight-binding and force-field codes.
Open tight-binding codes for large systems, such as KITE \cite{kite}, Kwant (with its \texttt{kpm} module) \cite{kwant}, pybinding \cite{pybinding} and TBPLaS \cite{tbplas}, use Chebyshev-based methods (KPM, or time propagation in TBPLaS) for orthogonal tight-binding models; I did not find an overlap matrix in their documentation or papers.
What is offered here is an open, tested implementation and a quantitative validation, and I want to be clear about what is old.
Chebyshev moments, KPM and Maxent for densities of states, impartial random vectors, locally truncated (exact) low-order moments and non-orthogonal KPM with PCG all come from the work of Silver, R\"oder, Voter, Kress, Drabold, Sankey and collaborators \cite{drabold1993,sankey1994,silver1994,silver1996,voter1996,roder1996,silver1997,roder1997}.
Silver and R\"oder \cite{silver1997}, in particular, fitted up to 1024 exact Chebyshev moments by maximum entropy for a tight-binding Hamiltonian of amorphous diamond with 16\,384 states; their prescription is described in Sec.~\ref{sec:recon}. Voter, Kress and Silver \cite{voter1996} obtain low-order moments from locally truncated Hamiltonians, which, as I understand it, is the same local computation as my exact low-order moments.
What I believe is new here is more limited: (i) using these exact moments up to order $K$ \emph{together with} sampled moments above $K$ in one estimator, with an argument for when it is unbiased and a measurement of when it is not useful (Secs.~\ref{sec:exact} and \ref{sec:small}); (ii) a Maxent solver with a moment-uncertainty floor and a staged (bootstrapped) climb in the number of moments, which I tested against exact spectra but \emph{not} against the Silver--R\"oder prescription; (iii) alternatives for the overlap inverse; and (iv) the open implementation, with chunked builders and matrix-free operators for $10^6$ atoms and more. I may have missed earlier descriptions of (i) and (ii).
Each section says what is established elsewhere and what was measured here.

Section~\ref{sec:methods} describes the methods; Sec.~\ref{sec:validation} compares with exact results; Sec.~\ref{sec:apps} gives four applications (a 2\,097\,152-atom crystal, the fullerene-to-graphene series, an amorphous Si model with DFTB and ACE, and a million-atom replica of a 216-atom cell); Sec.~\ref{sec:scaling} treats cost, memory and limits; Sec.~\ref{sec:recs} gives recommendations and Sec.~\ref{sec:conc} the conclusions and availability.

\section{Methods}
\label{sec:methods}

\subsection{Moments from random vectors}
\label{sec:moments}

Let $H$ be Hermitian with spectrum in $[E_{\min},E_{\max}]$, and $\tilde H=(H-b)/a$ with the interval widened by a margin $\epsilon w$ ($w=E_{\max}-E_{\min}$) on each side, $a=\tfrac12 w(1+2\epsilon)$, $b=\tfrac12(E_{\max}+E_{\min})$, so that the spectrum of $\tilde H$ lies in $(-1,1)$ even though the extreme eigenvalues are only estimated ($\epsilon=0.01$ by default).
The density per orbital is $\rho(E)=\frac{1}{a}\,\tilde\rho(x)$, $x=(E-b)/a$, where $E$ is the energy (the eigenvalue, for a general matrix), with Chebyshev moments
\begin{equation}
\mu_n=\frac1N\Tr\,T_n(\tilde H)=\int_{-1}^{1}\tilde\rho(x)T_n(x)\,dx .
\end{equation}
A random vector $\xi$ with $\mathbb E[\xi\xi^\dagger]=1$ gives $\mu_n\approx\frac1{N}\,\xi^\dagger T_n(\tilde H)\xi$ (for unit-variance entries), and averaging over $R$ vectors reduces the sampling error as $R^{-1/2}$; the error of each moment also falls as $N^{-1/2}$ (a bound is given below), so large systems need fewer vectors \cite{hutchinson,skilling,lin}.
\sdos\ draws Rademacher ($\pm1$) vectors for real and random phases for complex problems (Gaussian vectors are available; they have a larger variance, Table~\ref{tab:vec}).
For a Hermitian matrix $A$ with $\mathbb E[\xi\xi^\dagger]=1$ and independent entries, the variance of the estimate $\xi^\dagger A\xi$ of $\Tr A$ is $2\sum_{i\ne j}A_{ij}^2$ for Rademacher vectors (real symmetric $A$), $\sum_{i\ne j}|A_{ij}|^2$ for random phases and $2\|A\|_F^2$ for real Gaussian vectors; Rademacher vectors have the smallest variance among \emph{real} vectors with independent unit-variance entries \cite{hutchinson}; random phases halve the off-diagonal variance of a real symmetric $A$ at the price of complex arithmetic. With $A=T_n(\tilde H)$ and the factor $N^{-1}$ of the normalized moment, the standard deviation of $\mu_n$ for one Rademacher vector is bounded by $\sqrt{2/N}$, whatever the locality of $H$: $\|T_n(\tilde H)\|_2\le1$ gives $\|T_n(\tilde H)\|_F^2\le N$ and $\mathrm{sd}(\mu_n)=N^{-1}\bigl(2\sum_{i\ne j}[T_n]_{ij}^2\bigr)^{1/2}\le\sqrt{2/N}$. The covariance of the moments of different order and the cases where it matters are in Sec.~\ref{sec:small}. Variance-reduction schemes for stochastic trace estimation that I did not implement or test include probing \cite{tang} and low-rank deflation as in Hutch++ \cite{hutchpp}.
With $v_n=T_n(\tilde H)\xi$, the inner product $\langle u|v\rangle\equiv N^{-1}u^\dagger v$, the recursion $v_{n+1}=2\tilde Hv_n-v_{n-1}$ and the doubling identities
\begin{align}
\mu_{2n}&=2\langle v_n|v_n\rangle-\mu_0,\\ \mu_{2n+1}&=2\langle v_{n+1}|v_n\rangle-\mu_1,
\end{align}
give $M$ moments from $M/2$ matrix--vector products per vector.
For complex Hermitian $H$ all inner products are conjugated and everything else is unchanged; real arithmetic is used whenever the operator is real.
The operator may be an explicit sparse matrix (CSR) or any matrix-free linear operator; Hermiticity is checked (explicitly or by randomized probes for matrix-free operators), which matters for finite-difference Hessians that are Hermitian only to $O(\delta^2)$.

\paragraph*{Spectral bounds.}
The bounds are estimated with ARPACK (Lanczos)\footnote{The relative ARPACK tolerance is $10^{-3}$: the bound estimate then costs about 222 matrix--vector products for the 131\,072-orbital stencil, and the bounds move by less than 5~meV against a window of 20~eV.}, starting from a fixed vector (a Gaussian vector from a fixed internal seed, independent of the random vectors used for the moments). Two runs therefore give identical bounds, and the same bounds serve all moment vectors. The vectors of the moments themselves come from the user's \texttt{seed} and differ from run to run if none is given. Lanczos Ritz values lie \emph{inside} the spectrum, hence the margin $\epsilon$. For matrix-free operators built from finite differences of forces the estimate can fail badly (Sec.~\ref{sec:scaling}, item iii), and the package does not detect an over-wide window.
For explicit matrices Gershgorin bounds are available as a safe fallback.

\subsection{Exact low-order moments and impartial vectors}
\label{sec:exact}

Drabold and Sankey \cite{drabold1993} introduced \emph{impartial} vectors, chosen to weigh all eigenvectors equally, and computed low-order moments exactly; Voter, Kress and Silver \cite{voter1996} and Silver and R\"oder \cite{silver1997} evaluate low-order moments from locally truncated Hamiltonians. The substitution below uses that local computation.
The construction of Drabold and Sankey minimizes the penalty $P(\xi)=(\langle\xi|\xi\rangle-1)^2+(\langle\xi|H|\xi\rangle-\mu_1)^2+(\langle\xi|H^2|\xi\rangle-\mu_2)^2$ by conjugate gradients, with the exactly known $\mu_1=N^{-1}\Tr H$ and $\mu_2=N^{-1}\Tr H^2$ (both $O(N)$ from the matrix), so that the first three moments of $\xi$ are exact. This narrows the distribution of the band energy over the random vectors.
\emph{Only three moments are constrained} in the construction of Drabold and Sankey; constraining more makes the ensemble measurably biased (Sec.~\ref{sec:small}), so \sdos\ keeps three.
The recommended way to use exact low moments is \emph{substitution}: $\mu_n$ for $n\le K$ can be computed exactly in $O(N)$ because $[T_n(\tilde H)]_{ii}$ depends only on a neighbourhood of radius $\approx n/2$ hops of site $i$, and in every sample the sampled $\mu_n$ with $n\le K$ is replaced by the exact one (\texttt{exact\_moments=K}). The moments with $n>K$ remain the sampled estimates and the Chebyshev recursion is unchanged. The reconstruction uses one list, $\mu_0,\dots,\mu_K$ exact and $\mu_{K+1},\dots,\mu_{M-1}$ sampled, of one matrix.
Each $\mu_n$ is unbiased on its own, so every linear functional of the list (the KPM density, the integrated density, the band energy at a fixed Fermi level) is unbiased, and the variance falls sharply (Table~\ref{tab:vec}). The list is not exactly the moment list of a positive density; this does not matter for KPM, and with Maxent it is absorbed by the sampled uncertainties of $R\ge2$ vectors (Sec.~\ref{sec:small}), so with a single vector, which has no variance estimate, \sdos\ uses $K=0$.
The default is $K=4$ when $A$ is an explicit sparse matrix and $B$ is absent or diagonal, and $K=0$ otherwise, because the exact moments need the matrix elements. The cost per site grows polynomially with $K$, not with $N$.
Table~\ref{tab:vec} measures all options on one problem: at equal cost every variance-reduction option beats plain Rademacher vectors by a factor of about 35--50, with $K=8$--16 comparable (the quoted uncertainty is 30\%).

\begin{table}[t]\caption{Random vectors and exact low moments, band energy of DTW4096 (Kwon, $M=100$, 12 seeds with $R=2$ each; std is the standard deviation of an estimate with $R=2$ vectors, $\sqrt2$ times larger per vector). Efficiency = (variance $\times$ time)$^{-1}$ relative to plain Rademacher vectors; the uncertainty of each entry is about 30\%.}\label{tab:vec}\begin{ruledtabular}\begin{tabular}{lrrr}
vectors & std (meV/at.) & time (s) & efficiency \\ \colrule
Rad. & 192.8 & 0.23 & 1.0$\times$ \\
Gauss. & 315.4 & 0.23 & 0.4$\times$ \\
impartial (3) & 25.6 & 0.35 & 36.7$\times$ \\
Rad. + exact K=4 & 11.2 & 1.75 & 38.0$\times$ \\
Rad. + exact K=8 & 4.4 & 8.99 & 47.3$\times$ \\
Rad. + exact K=16 & 1.4 & 79.84 & 52.2$\times$ \\
\end{tabular}\end{ruledtabular}\end{table}

\subsection{Reconstruction: KPM and maximum entropy}
\label{sec:recon}

\paragraph*{KPM.}
The default reconstruction is the damped series $\tilde\rho(x)=\frac{1}{\pi\sqrt{1-x^2}}\bigl[g_0\mu_0+2\sum_{n\ge1}g_n\mu_nT_n(x)\bigr]$ with the Jackson kernel $g_n$ \cite{jackson,kpm} (positive, energy resolution $\approx\pi a/M$ throughout this report, taken as the Gaussian $\sigma$ of the kernel, so that the full width at half maximum is about $2.355\,\pi a/M$; the Fej\'er sums \cite{fejer,courant} are the simplest damping with a positive kernel); the Lorentz kernel is available for Green-function-like uses.
The density is evaluated on the Chebyshev--Gauss nodes $x_k=\cos\theta_k$, $\theta_k=\pi(k-\tfrac12)/n_{\rm pts}$ with $n_{\rm pts}=\max(2M,1000)$ by default; the integrated density is the analytic integral of the damped series, and the band energy and Fermi level are obtained from it on a uniform grid of 20\,000 points in $\theta$ (trapezoid rule in $\theta$).
KPM is linear in the moments and never fails to converge. Beyond the resolution at which the sampling noise dominates, more moments only sharpen speckle. For a finite system with $N$ levels, a resolution finer than the mean level spacing resolves individual levels.

\paragraph*{Maximum entropy.}
Maxent \cite{mead,turek,drabold1993,silver1997} writes $\tilde\rho(x)=m(x)\exp[\sum_n\lambda_nT_n(x)]$ with a default model $m$ and finds $\lambda$ by Newton's method on the convex dual
\begin{equation}
\mathcal D(\lambda)=\int m\,e^{\lambda\cdot T}dx-\lambda\cdot\mu+\tfrac12\sum_n\sigma_n^2\lambda_n^2 ,
\end{equation}
whose integrals are evaluated by the midpoint rule on a uniform grid in $\theta=\arccos x$ of $\max(8192,64M)$ points (so that $dx=\sin\theta\,d\theta$ and a flat model is $\tfrac12\sin\theta$); the integrated density is the cumulative midpoint sum, and the band energy follows from a trapezoid rule in energy over its increments. The dual has gradient equal to the moment residual plus $\sigma_j^2\lambda_j$ and Hessian $\int m\,e^{\lambda\cdot T}T_jT_k\,dx+\mathrm{diag}(\sigma_j^2)$. Since $T_jT_k=\tfrac12(T_{j+k}+T_{|j-k|})$, the integral is a function of the Chebyshev moments of the current iterate $m\,e^{\lambda\cdot T}$. Here $\sigma_n$ are the standard errors of the sampled moments (zero for exact moments), and the solver uses an eigenvalue-floored Newton step with backtracking.
The default model is uniform (\texttt{flat}); the Chebyshev weight, the Jackson-KPM density and a KPM model at fewer moments gave no measurable difference in accuracy or conditioning.
Two practical points follow from my tests.
First, with a spectral window (the interval $[E_{\min},E_{\max}]$ used for the scaling) much wider than the support (the interval that contains the eigenvalues; a poor bound), the multipliers become very large (up to $10^4$--$10^5$ for exact moments) although the density remains accurate; stochastic moments with their errors tolerate even a factor-two window.
Second, Maxent with \emph{exact} moments of a density with singularities (log van Hove peaks, Dirac cusps) is a badly conditioned problem. This is an empirical observation; I have no proof. Silver and R\"oder \cite{silver1997} fitted 1024 exact moments for a 16\,384-state amorphous-diamond Hamiltonian; they fitted Gibbs-damped moments (the positive-kernel factors $g_m^{M\times K}$) with the kernel-polynomial density as the default model, evaluated the Fourier integrals on a fine grid of $M\times K\times I$ pixels (they use $K=8$, $I=4$), used moment uncertainties at the numerical precision for exact data and continued in the regularization parameter $\alpha$ (an outer loop that decreases $\alpha$ from a large value) rather than in the number of moments, none of which I tested, so the non-convergence reported here refers to the solver described in this section (undamped moments, flat model), not to theirs. The exact moments of a density with van Hove peaks and band edges lie close to the boundary of the set that an exponential $\exp(\sum_n\lambda_nT_n)$ can reproduce, the multipliers grow as $M$ increases, and the Newton iteration stops converging: for DTW4096 it did not converge in 300 steps for $M\ge100$ (Table~\ref{tab:acc}). The behaviour depends on the density: for C$_{3840}$ with $M=150$ exact moments the plain solver does converge (403 steps without the floor, 27 with it; Sec.~\ref{sec:app_fu}). With sampled moments the sampling error regularizes the problem.
The same regularization can be applied to exact moments: \sdos\ assigns every moment except $\mu_0$ a minimum uncertainty $\sigma_n\ge s_0$ with $s_0=10^{-4}$ by default (\texttt{sigma\_floor}); the moments are normalized, $|\mu_n|\le1$, so this is an absolute scale.
The dual then has a finite solution and Newton's method converges in 9--24 steps at all $M$ tested (Table~\ref{tab:acc}). For stochastic moments of a 16\,384-orbital system with $R=4$ and $K=0$ the sampled errors ($\sigma_n$ between $7\times10^{-4}$ and $4\times10^{-3}$ for $n>K$) exceed the floor, which then changes nothing (identical results for $M=100$, 170 and 300). The floor does act on the exact moments $n\le K$ ($\sigma_n=0$ there; $K=4$ by default), and it is essential for $R=1$, where there is no variance estimate (Table~\ref{tab:rv}).
The floor is a bias of order $s_0$ in the moments, which was found to be $\le0.02$~meV/atom in the band energy at $s_0=10^{-4}$ and $0.2$~meV/atom at $10^{-3}$; \texttt{sigma\_floor=0} restores the exact-moment problem.
Whether $10^{-4}$ is small enough depends on the quantity. A shift $\delta\mu_n$ of the moments changes an integrated quantity $\int f\tilde\rho\,dx$ by $\sum_nf_n\delta\mu_n$, with $f_n$ the Chebyshev coefficients of $f$; for the band energy $f=(ax+b)\Theta(x_F-x)$ and $|f_n|\sim a/n$, so the worst case, with every moment displaced coherently by $s_0$, is $\sim s_0\,a\ln M$ (a few meV per orbital for $a=10$~eV), and a random-sign shift is $\sim1.3\,s_0a$. These are bounds on the bias of the constraints, not on the fitted density, whose error was measured to be far smaller (above). A quantity that depends on fine structure at high order (a feature narrower than $\pi a/M$, a sharp edge, a spectral weight below $10^{-4}$) or that needs a precision better than $\sim s_0a$ may be affected: reduce \texttt{sigma\_floor}, accepting slower convergence and, in the limit $s_0=0$, the non-convergence documented above.
In practice: with exact moments the default floor gives the exact-moment Maxent density, to a band-energy error of $0.02$~meV/atom or less, in a few tens of Newton steps where the plain solver fails; with sampled moments, $K=0$ and $R\ge4$, the floor is inactive. More random vectors do not help the exact-moment problem, because the exact moments do not depend on $R$. For sampled moments they lower the sampling error and with it the pointwise error (Table~\ref{tab:rv}), but convergence is not the issue there: Newton converges in 12--15 steps for $R\ge4$.

\paragraph*{Bootstrapped Maxent.}
\texttt{maxent\_bootstrap} climbs in the number of moments: it solves for $M=16,32,64,\dots$ in turn, starts each stage from the previous multipliers, and after each stage restricts the integration support of the dual to the interval where the density found at that stage exceeds $10^{-4}$ of its maximum, padded by two resolution lengths $\pi/M$ (in the scaled variable $x$).
No states are allowed outside the edges, so the multipliers need not build huge exponentials to suppress the density in the empty margin of the window.
The support is a modelling statement, checked by the residual of the moments; a stage that fails to converge is widened and retried, and if it still fails the last converged stage is returned (with \texttt{converged=False} and the number of moments used).
For graphene (Xu Hamiltonian, exact $k$-sum on a $240\times240$ mesh) with exact moments, $M=400$, the bootstrap converges in 29~s without the uncertainty floor, where the plain solver does not converge in 300 steps (46~s; $L_1$ distance to the exact spectrum broadened by 0.05~eV: $0.0446$ for the bootstrap, and $0.0449$ for the last iterate of the plain solver, which is not a converged result). With the default floor both converge, the bootstrap in 2.0~s and the plain solver in 3.8~s (31 steps), at $L_1=0.0444$ (timings on 2 threads, approximate). The support barely shrinks for graphene, because the band fills its window, so here the gain comes from the warm starts.

\paragraph*{Integrated quantities.}
The integrated density $N(E)$, the Fermi level for a given electron count, and the band energy $E_{\rm BS}=2\sum_{E_i<E_F}E_i$ follow from the reconstruction (\texttt{fermi\_and\_band\_energy}), with jackknife error bars for Maxent (the spread of the result when each random vector is left out in turn).
In a gap the integrated density is flat, so $E_F$ is not determined by the moments to better than the gap: \sdos\ reports the interval over which $N(E)$ equals the electron count within noise (\texttt{fermi\_interval}) and warns; the band energy is nearly independent of where $E_F$ is placed in a true gap.
Exact HOMO and LUMO levels are obtained separately (Sec.~\ref{sec:windows}).

\subsection{Generalized problems}
\label{sec:generalized}

For $A\psi=\lambda B\psi$, $Y=B^{-1}A$ is self-adjoint in the $B$-inner product and similar to $B^{-1/2}AB^{-1/2}$, so $\Tr\,T_m(\tilde Y)=\Tr\,T_m(\widetilde{B^{-1/2}AB^{-1/2}})$ and $\xi^\dagger T_m(\tilde Y)\xi$ is an unbiased estimator for any $\xi$ with $\mathbb E[\xi\xi^\dagger]=1$ \cite{roder1996}.
$Y$ is not Hermitian in the Euclidean product, so the doubling identities do not apply: one application of $B^{-1}$ per moment is needed, and the cost per moment is that of a solve.
Spectral bounds of $Y$ use a Lanczos process in the $B$-inner product.
\sdos\ offers the following ways of applying $B^{-1}$, all verified to give the same moments to tolerance (Table~\ref{tab:dd}):
(i) PCG to a relative residual $10^{-12}$ with block-Jacobi (default), incomplete Cholesky IC(0) with a Manteuffel shift \cite{meijerink,manteuffel} (the shift needed is reported, since the overlap of a minimal basis is not an \mbox{M-matrix}), or algebraic multigrid;
(ii) a fixed polynomial $p(\tilde B)\approx\tilde B^{-1}$ on the diagonally scaled $\tilde B=D^{-1/2}BD^{-1/2}$ (the scaled problem $\tilde Y=D^{1/2}YD^{-1/2}$ is similar to $Y$), of degree $\approx\tfrac12\sqrt\kappa\ln(2/\varepsilon)$ from the condition number $\kappa$. In detail: the extreme eigenvalues of $\tilde B$ are found by ARPACK (tolerance $10^{-8}$) and widened by 2\% on each side to guard against bounds that are too tight; $p$ is the Chebyshev interpolant of $1/t$ on that interval; its degree starts from $\ln(2/\varepsilon)/\ln(1/q)$ with $q=(\sqrt\kappa-1)/(\sqrt\kappa+1)$ and is raised until $\max|1-t\,p(t)|\le\varepsilon$ on a grid of 4001 points ($\varepsilon=10^{-12}$ by default, \texttt{cheb\_tol}); $p(\tilde B)$ is applied to blocks of vectors by the Clenshaw recurrence, one product with $\tilde B$ per degree, and a check on a random vector confirms the tolerance. The inverse is then replaced by a matrix polynomial, so the recursion is applied to one fixed operator and needs no inner solves; the error of the polynomial is a known, controlled approximation.
For a diagonal $B$ (masses) $B$ is absorbed into $A$ once and no solves are needed.
For the DFTB overlap of silicon ($\kappa(\tilde B)=3.85$) the polynomial has degree 26 and PCG with block-Jacobi needs about 24 iterations per solve (Table~\ref{tab:dd}); the iteration count is independent of the system size (Sec.~\ref{sec:scaling}).

\subsection{Eigenvectors in spectral windows, exact gaps and projections}
\label{sec:windows}

\paragraph*{Windows.}
\texttt{window\_eigs} returns the eigenpairs with $\lambda$ in a window together with their inverse participation ratio $\mathrm{IPR}=\sum_s w_s^2$ (site weights $w_s$ from the Mulliken populations, see Projections below), by shift-invert (sparse $LDL^\dagger$ of $A-\tau B$, ARPACK) or by Chebyshev-filtered subspace iteration \cite{zhou}.
\emph{Shift-invert is not $O(N)$}: the fill of a sparse factorization grows faster than linearly in three dimensions; the filtered iteration (a Chebyshev polynomial filter) is the scalable fallback. A filter of a given degree separates the states in the window from the rest of the spectrum only at distances from the window edges larger than about $2\pi a/\mathrm{degree}$, where $a$ is the half-width of the spectrum.
\paragraph*{Exact HOMO--LUMO gaps.}
By Sylvester's law of inertia, the number of eigenvalues below $\tau$ equals the number of negative pivots of $A-\tau B$; \texttt{count\_states\_below} evaluates it from a symmetric factorization without pivoting, and \texttt{homo\_lumo} finds HOMO and LUMO by two bisections, with no eigensolver and no assumption on the gap size or on degeneracy.
Because of the fill, this is more expensive than the density of states above $\sim2\times10^4$ orbitals (Sec.~\ref{sec:scaling}), so the exact gap is a tool for moderate sizes.

\paragraph*{Projections.}
\label{sec:proj}
For $A\psi=\lambda B\psi$ with eigenvectors normalized $\psi^\dagger B\psi=1$, the weight of orbital $i$ in state $n$ is the \emph{Mulliken} weight $w_{in}=\mathrm{Re}\,\psi_{in}^*(B\psi_n)_i$ \cite{mulliken} (equal to $|\psi_{in}|^2$ for $B=1$).
For overlapping orbitals $w_{in}$ is \emph{not} a probability: it can be negative or exceed one (the sum over $i$ is one, $\sum_iw_{in}=\psi_n^\dagger B\psi_n$), so a local or projected density built from it is a signed quantity, not a positive local spectral density, and it can be negative at some energies for orbitals that overlap strongly with their neighbours. This is a property of the Mulliken partition, as for Mulliken charges; the total density is unaffected. KPM handles signed densities; Maxent assumes positivity and is not used for projections. A positive alternative is the L\"owdin partition, which needs $B^{1/2}$ and is not provided.
The order of the operators matters. With $Y=B^{-1}A$ and $\sum_n\psi_n\psi_n^\dagger=B^{-1}$ the spectral projector is $\sum_n\psi_n\psi_n^\dagger\,\delta(E-\lambda_n)=\delta(E-Y)B^{-1}$. Its diagonal element $[\delta(E-Y)B^{-1}]_{ii}=\sum_n|\psi_{in}|^2\delta(E-\lambda_n)$ integrates to $(B^{-1})_{ii}$, not to one, so it is \emph{not} the Mulliken density. With the Mulliken weight $w_{in}$ defined above,
\begin{multline}
\mathrm{LDOS}_i(E)=\mathrm{Re}\sum_n\psi_{in}^*(B\psi_n)_i\,\delta(E-\lambda_n)\\=\mathrm{Re}\,[B\,\delta(E-Y)\,B^{-1}]_{ii}=\mathrm{Re}\,[\delta(E-Y^\dagger)]_{ii},
\end{multline}
because $BYB^{-1}=AB^{-1}=Y^\dagger$ ($Y^\dagger$ the Euclidean adjoint of $Y$) and $f(Y)B^{-1}=B^{-1}f(AB^{-1})$ (the matrix $B$ does not commute through the function; it conjugates the argument). Its Chebyshev moments are the diagonal elements $[T_n(\tilde Y^\dagger)]_{ii}=e_i^\dagger T_n(\tilde Y^\dagger)e_i$ of the adjoint operator, which is applied to a vector as $Y^\dagger v=A(B^{-1}v)$: first the inverse of $B$, then $A$ (the reverse of $Yv=B^{-1}(Av)$ used for the total density). The code applies exactly this operator (\texttt{GeneralizedOperator.apply\_adjoint}), and a unit test with dense $B$, real and complex, compares the resulting moments of single orbitals with the Mulliken weights $\mathrm{Re}\,\psi_{in}^*(B\psi_n)_i$ of the dense generalized eigenproblem to $10^{-7}$.
Then $\mathrm{PDOS}_g=\sum_{i\in g}\mathrm{LDOS}_i$.
Sum rules hold exactly: $\sum_i\mathrm{LDOS}_i=N\rho$, each LDOS integrates to one, and $\mathrm{PDOS}_g$ integrates to $n_g$, the number of orbitals in the group.
Two estimators are provided (projected random vectors were used for local densities by Sankey, Drabold and Gibson \cite{sankey1994}): ``shared'' (one random vector is propagated once and every group reads its own partial inner product, all groups for the price of one DOS, at the cost of extra noise for small groups) and ``separate'' (a vector supported on the group), and \texttt{ldos} propagates unit vectors for a deterministic result.

\subsection{Vibrations}
\label{sec:vib}

For a force-constant matrix $\Phi$ and masses $m_i$, the dynamical matrix $D_{ij}=\Phi_{ij}/\sqrt{m_im_j}$ has eigenvalues $\omega^2$.
\texttt{vdos} works with $\lambda=\omega^2$ and reports $g(\omega)=2\omega\rho(\omega^2)$; negative eigenvalues (instabilities) are reported, not clipped, with the weight below the resolution.
The three rigid translations are known eigenvalues at $\lambda=0$ and can be removed exactly from the moments (\texttt{zero\_modes=3}), which matters for the entropy and the low-temperature heat capacity because the kernel would smear their weight into small positive $\omega$.
\texttt{harmonic\_thermo} gives the zero-point energy, $U$, $F$, $S$ and $C_v$ per atom and warns below the temperature where the lowest frequencies are unresolved.
\paragraph*{Matrix-free Hessians.}
For large models \texttt{HessianOperator} applies $\Phi u=-[F(x_0+su)-F(x_0-su)]/2s$ with any force provider (LAMMPS through its Python module). In this case a product costs two force calls and translations are projected out.
Units and conventions: positions in \AA, forces in eV/\AA, masses in atomic mass units (the LAMMPS \texttt{metal} units), so $\Phi$ is in eV/\AA$^2$ and $D$ in eV\,\AA$^{-2}$\,u$^{-1}$, with $\omega^2=\lambda$ and $\hbar\omega=64.654~\mathrm{meV}\times\sqrt{\lambda/[\mathrm{eV\,\AA^{-2}\,u^{-1}}]}$ in the thermodynamics (the factor is the argument \texttt{energy\_per\_sqrt\_lambda} of \texttt{harmonic\_thermo}, for other unit systems). The cell is periodic as defined by the force provider (LAMMPS applies the minimum image and its own neighbour lists), and the Hessian is that of the structure at $\Gamma$. For a vector $v$ the displacement is $\delta x=s\,m^{-1/2}v$ (components $v/\sqrt{m_i}$), where the scalar $s$ is chosen such that the rms of the Cartesian components of $\delta x$, $\bigl(\tfrac1{3N}\sum|\delta x_{i\alpha}|^2\bigr)^{1/2}$, equals $10^{-3}$~\AA\ regardless of $N$ and of the norm of $v$ (so the rms displacement per atom, as a vector, is $\sqrt3\times10^{-3}$~\AA); the central difference then gives $\Phi\,\delta x$ with an $O(\delta^2)$ error and the result is multiplied by $m^{-1/2}$ again to give $Dv$.
No Hessian is stored.

\subsection{Interfaces and implementation}
\label{sec:impl}

\sdos\ is a pure-Python package (NumPy \cite{numpy}, SciPy \cite{scipy}; Python~3.14 in the present tests); everything else is optional.
\emph{Tight-binding models}: the Hamiltonian of Kwon \emph{et al.}\ \cite{kwon}, an empirical, orthogonal $sp^3$ tight-binding total-energy model for silicon (a band-structure term plus a repulsive pair potential; \sdos\ uses only its Hamiltonian matrix, with four orbitals per atom), with Bloch twists (complex Hermitian matrices), a chunked assembler for $10^6$ atoms and a matrix-free periodic stencil for crystals; and the Xu--Wang--Chan--Ho carbon model \cite{xu}.
\emph{DFTB} Slater--Koster (\texttt{.skf}) files, read from a user-supplied directory (the \texttt{pbc-0-3} set \cite{pbc03} is distributed under CC BY-SA~4.0 and requires citation of Refs.~\cite{sieck2000,sieck2003}; it is not bundled), non-self-consistent \cite{porezag,elstner} (zeroth-order Hamiltonian $H_0$ and overlap $S$).
\emph{SIESTA} $H$ and $S$ through \texttt{sisl} \cite{sisl,siesta2002,siesta2020}, and \emph{LAMMPS} force fields \cite{lammps} through the matrix-free Hessian.
A command-line tool (\texttt{sdos}) covers the common cases.
The test suite has 186 tests (run with OpenMP/BLAS capped at 4 threads), including the comparisons with exact diagonalization listed in Sec.~\ref{sec:validation}, a regression test against the original single-file code and tests that the documented examples run.

\paragraph*{Input, result files and analysis.}
The simplest matrix file is plain text with one entry per line, \texttt{i j a\_ij} (a fourth column is the imaginary part); one or both triangles, 0- or 1-based indices and the storage are detected, and the other triangle is generated by conjugation (Matrix Market and SciPy \texttt{.npz} are also read).
The density-of-states stage needs only the matrix and produces a density $\rho(E)$ per orbital, normalized to one.
A result written with \texttt{sdos H.dat -o result.npz} (here \texttt{H.dat} is a plain-text list of the nonzero matrix elements; or \texttt{res.save}) is a self-describing file: $\rho$, its error bars and the integrated density, the averaged and per-sample Chebyshev moments, the spectral window, the method and its settings, and for vibrations $\omega$ and $g(\omega)$.
\texttt{sdos analyze result.npz} (or \texttt{sdos.load}) then works from the file alone, without the matrix.
Whatever needs information from outside the matrix enters only here: for electrons the number of electrons (given directly or as the sum of the valence of each atom, from a file) and the spin degeneracy give $E_F$ and the band energy; for vibrations a temperature grid gives the zero-point energy and $U$, $F$, $S$, $C_v$ (the masses and the three zero modes change the moments and are therefore applied at the density-of-states stage).
The analysis can also rebuild the density from the first $M'$ saved moments with KPM or Maxent, which costs no matrix products.
Saving and loading reproduces the arrays exactly, and $E_F$ and $E_{\rm BS}$ from a loaded file agree with the in-memory values to $10^{-12}$ for both methods (tests); reconstruction from a loaded vibrational result is not supported, and the generalized problem was not tested through the file.

\paragraph*{The DFTB reader.}
Reading the DFTB format correctly required three conventions that I did not find stated in the documentation available to me (the DFTB+ manual; I could not retrieve the separate Slater--Koster format description, and some of them may be documented there), which I recovered by comparing whole matrices with DFTB+ \cite{dftbplus} (Sec.~\ref{sec:val_dftb}): a table announced as \texttt{ngrid} rows contains $\texttt{ngrid}-1$, row $j$ sits at $r=j\,\Delta r$ ($r=0$ is the omitted row) and the first 19 rows of the silicon file are placeholders for $r<0.4$~Bohr; the $p$ orbitals are ordered $(p_y,p_z,p_x)$; and each integral is continued past the end of its table by a quintic that reaches zero, with vanishing first and second derivatives, one Bohr later and matches value, slope and curvature of the table's last row.
The non-orthogonal pair convention follows from the file name: in \texttt{A-B.skf} the first orbital of each integral sits on A, so a ($p$ on A, $s$ on B) block uses the $B$--$A$ file.

\section{Validation}
\label{sec:validation}

The electronic examples use the Kwon Hamiltonian \cite{kwon}, an empirical, orthogonal $sp^3$ tight-binding total-energy model of silicon; its spectra are those of that model, so agreement with an exact spectrum below tests the method, not the model. Each test below is compared with exact diagonalization or with an independent code; the benchmarks use DTW4096, a 4096-atom model of amorphous silicon from the Djordjevic--Thorpe--Wooten family \cite{www,dtw} (relaxed at zero pressure with ACE, cell $43.6\times43.7\times43.9$~\AA$^3$), at $\Gamma$ unless stated, since the cells are large and the $k$-point dependence of an amorphous Si cell this size is weak.

\paragraph*{Conventions.}
Unless stated otherwise, $L_1=\int|\rho_{\rm est}-\rho_{\rm ref}|\,dE$ over the whole spectrum with both densities normalized to one, where $\rho_{\rm ref}$ is the histogram of the exact eigenvalues broadened by a Gaussian of the width given in each case (0.05~eV for the tests with exact moments, 0.1~eV for the stochastic ones; a narrower reference penalizes the sharper Maxent curves, a wider one the broader KPM curves). Band energies are for the electron count of the neutral system, in meV per atom, and ``error'' is estimate minus exact. A sampling error quoted as $\pm$ is the standard error over the $R$ vectors unless it is marked $2\sigma$.
Table~\ref{tab:valsum} lists every feature that I checked, with the system, the reference, the quantity and the observed error; the last rows list what was not checked or not reproduced.

\begin{table*}
\caption{Summary of the validation. ``Dense'' is a dense diagonalization of the same matrix. DTW4096: 4096-atom model of amorphous silicon with the Kwon tight-binding Hamiltonian (16\,384 orbitals) or DFTB; ``Kwon TB'' is the empirical, orthogonal tight-binding total-energy Hamiltonian of Kwon \emph{et al.}\ \cite{kwon}; its spectra are those of this model and not of a first-principles calculation.}
\label{tab:valsum}
\begin{ruledtabular}
\begin{tabular}{p{2.9cm}p{3.1cm}p{2.6cm}p{2.9cm}p{4.6cm}}
Feature & System & Reference & Quantity & Observed error \\ \colrule
Exact local moments ($K\le40$) & chain, dimerized chain, Kwon TB, 64--864 orbitals & dense moments & $\max_n|\Delta\mu_n|$ & $\le5\times10^{-15}$ \\
KPM truncation (Jackson) & DTW4096, Kwon TB & dense, exact moments & $E_{\rm BS}$; $L_1$ (0.05~eV) & $+0.09$ to $+0.30$~meV/atom for $M=100$--400; $L_1=3.7\times10^{-3}$ at $M=400$ \\
Maxent, exact moments, floor $10^{-4}$ & DTW4096, Kwon TB & dense, exact moments & $E_{\rm BS}$; $L_1$ (0.05~eV) & $-0.017$ to $+0.02$~meV/atom for $M=100$--400; $L_1=0.007$--$0.012$ \\
Sampling error & DTW4096, Kwon TB, $M=100$ & dense & s.d.\ of $E_{\rm BS}$ over seeds & $\approx270$~meV/atom per Rademacher vector (193 at $R=2$), $\propto R^{-1/2}$; $\approx16$ per vector (11 at $R=2$) with $K=4$ (Table~\ref{tab:vec}) \\
Substitution and impartial vectors at small $N$ & chain, dimerized chain, Kwon TB, Anderson, $N=32$--4000 & dense & RMSE of $E_{\rm BS}$ ($M=100$, $R=4$) & $0.04$--$0.10$ ($K=4$) and $0.004$--$0.014$ ($K=16$) of plain vectors (Sec.~\ref{sec:small}) \\
Complex Hermitian ($\mathbf k$ twist) & Kwon 64 atoms, DFTB 216 atoms & union of primitive-cell bands & eigenvalues; Hermiticity & $10^{-9}$, $10^{-8}$~eV; $10^{-12}$ \\
Matrix-free stencil & Kwon TB, $16^3$ cells (131\,072 orbitals) & exact $\mathbf k$-sum & $L_1$ (0.1~eV), $M=170$, $R=6$ & KPM 0.036, Maxent 0.047; Hermiticity $1.8\times10^{-17}$ \\
DFTB matrices & 64-atom Si cell, Si--H cluster & DFTB+ \texttt{WriteRealHS} & matrix elements; eigenvalues & $H$: $7.9\times10^{-7}$, $S$: $4.3\times10^{-8}$~eV; eigenvalues $3\times10^{-6}$~eV \\
Generalized, three $B^{-1}$ modes & DTW4096, DFTB, $\Gamma$ & dense generalized & $E_{\rm BS}$; $L_1$ (0.1~eV) & identical moments in the three modes; sampling error rms 110~meV/atom at $R=6$, 80 at $R=24$ (two seeds), $-0.3\pm7.7$ at $R=96$ (eight seeds; moment $z$-scores rms $1.00$, mean $0.006\pm0.036$); $L_1=0.024$ (Table~\ref{tab:dd}) \\
SIESTA matrices; inertia count & C$_{240}$, C$_{540}$ (960, 2160 orbitals) & dense $\mathrm{eigh}(H,S)$ & HOMO--LUMO gap & $2\times10^{-10}$, $4\times10^{-11}$~eV \\
Matrix-free Hessian & Si 512 atoms, ACE & LAMMPS \texttt{dynamical\_matrix} & matrix, relative & $2.4\times10^{-5}$ \\
Harmonic thermodynamics & DTW4096, ACE and SW & dense dynamical matrix & ZPE; $C_v$ (300~K); $C_v$ (100~K) & ACE: $-0.66\%$, $+0.27\%$, $+1.4\%$; SW: $-0.54\%$, $+0.29\%$, $+0.92\%$ (Table~\ref{tab:ph}) \\
Shift-invert window & DTW4096 & dense & three levels in the gap & $7\times10^{-14}$~eV \\
$17^3$ replica of R10, Kwon TB & 4\,244\,832 orbitals & exact Bloch spectra & max error of $N(E)$ & Maxent 0.04\%, KPM 0.08\% of $N$ (Sec.~\ref{sec:app_r10}) \\
$17^3$ replica of R10, ACE & 3\,183\,624 degrees of freedom & exact Bloch spectra & ZPE; low-frequency mode count & ZPE 0.03--0.07\%; 0.5\% of the modes below 9~meV at $M=600$ \\
2\,097\,152-atom crystal & Kwon TB, $64^3$ cells, matrix-free & exact $\mathbf k$-sum of the 8-atom cell & band edges; gap & edges within resolution; Maxent follows the exact edges (0.46, 1.24~eV); no $L_1$ computed \\
\colrule
Not reproduced & C$_{3840}$ gap & Ref.~\cite{ssc96} Fig.~3 & gap & $0.284$ against $0.10$~eV (open) \\
Not checked & complex generalized at $\mathbf k$; Mulliken weights and IPR; $10^6$-atom runs & -- & -- & only sum rules or an internal comparison (Sec.~\ref{sec:scaling}, item vii) \\
\end{tabular}
\end{ruledtabular}
\end{table*}

\subsection{Unit tests}

The package is tested against the following, all of which pass: (1) a cubic Anderson model; (2) the same with Peierls phases (complex Hermitian); (3) a 216-atom diamond-structure silicon cell with the Kwon Hamiltonian; (4) a random sparse $A$ with a sparse positive definite $B$ whose condition number is tuned from $\approx10$ to $\approx10^4$, real and complex; (5) a Keating \cite{keating} dynamical matrix with unequal masses in both forms (mass-weighted, and generalized with the mass matrix); (6) PCG (block-Jacobi) and the Chebyshev polynomial give the same moments (the other preconditioners, IC(0) and AMG, are exercised in a further test and compared on DTW4096 in Table~\ref{tab:dd}); (7) regression against the original code.
Beyond these: Maxent robustness (large windows, default models), the Mulliken sum rules, local exact moments against the dense matrix, the inertia counts, and the interfaces (Secs.~\ref{sec:val_cplx}--\ref{sec:val_vib}).

\subsection{Accuracy against the number of moments and vectors}
\label{sec:val_acc}

\begin{figure*}
\includegraphics[width=\textwidth]{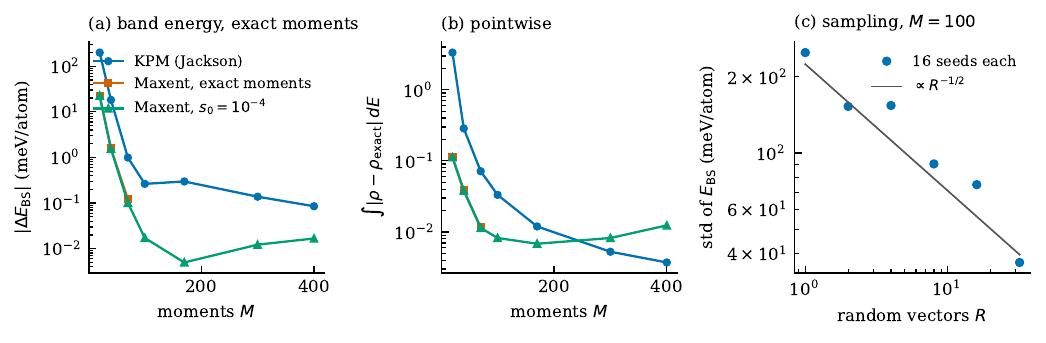}
\caption{Accuracy on DTW4096 (Kwon TB, $\Gamma$). (a) Band-energy error from exact moments: KPM (Jackson), Maxent with exact-moment constraints (converged cases only) and Maxent with the uncertainty floor $s_0=10^{-4}$. (b) Pointwise $L_1$ distance to the exact spectrum broadened by 0.05~eV. Maxent with exact-moment constraints is shown only where the solver converged. (c) Standard deviation of the band energy over 16 seeds at $M=100$ against the number of random vectors $R$ (plain Rademacher vectors, no exact-moment substitution), with the $R^{-1/2}$ law.}
\label{fig:acc}
\end{figure*}

To separate truncation from sampling I compute the moments \emph{exactly} from the dense spectrum of DTW4096 (Kwon TB, 16\,384 orbitals) and reconstruct (Table~\ref{tab:acc}, Fig.~\ref{fig:acc}a,b).
KPM with the Jackson kernel converges steadily ($L_1$ error $3.7\times10^{-3}$ at $M=400$); Maxent with exact-moment constraints is more accurate for the same moments up to $M=70$, the last size at which it converged: at $M=70$ its band-energy error is $0.12$~meV/atom against $1.0$ and its $L_1$ distance is six times smaller, at $M=40$ $1.6$ against $-18$~meV/atom.
From $M=100$ the plain Maxent solver no longer converges with exact moments of this singular density; how many exact moments the plain solver manages depends on the problem (for some it works to several hundred); those results are not reported (dashes in Table~\ref{tab:acc}, no points for ``Maxent, exact moments'' in Fig.~\ref{fig:acc}).
With the uncertainty floor $s_0=10^{-4}$ (Sec.~\ref{sec:recon}) every case converges: the band-energy errors are $+0.02$, $-0.005$, $-0.012$ and $-0.017$~meV/atom at $M=100$, 170, 300 and 400, 5--60 times smaller than KPM's ($+0.09$ to $+0.30$~meV/atom), and the $L_1$ distance to the 0.05~eV-broadened exact spectrum is 0.007--0.012.
The $L_1$ distance of Maxent $s_0$ is smallest at $M=170$ ($0.0068$) and then \emph{rises} ($0.0082$ and $0.0123$ at $M=300$ and 400; Fig.~\ref{fig:acc}b), while KPM's keeps falling and is smaller at $M=300$ and 400 ($0.0053$ and $0.0037$); the rise is not caused by the floor: at $M=400$ the $L_1$ distance is $0.0117$, $0.0123$ and $0.0128$ for $s_0=10^{-3}$, $10^{-4}$ and $10^{-5}$ ($s_0=10^{-6}$ did not converge in 300 steps, so I give no number), and at $M=300$ it is $0.0082$--$0.0083$ for $s_0=10^{-5}$ to $10^{-3}$ ($0.0095$ for $10^{-6}$, 285 steps; script \texttt{bench\_accuracy\_floor.py}, the setup of Table~\ref{tab:acc}). I attribute it to the ill-conditioning of the exact-moment problem for a singular density (Sec.~\ref{sec:recon}) and in part to the test, because a sharp Maxent curve is compared with a spectrum broadened by 0.05~eV, so for resolving fine structure KPM remains the choice and for integrated quantities Maxent is better here.
With sampled moments the sampling error dominates (Fig.~\ref{fig:acc}c). Table~\ref{tab:rv} shows what more random vectors do to Maxent with sampled moments: the $L_1$ error falls with $R$ (roughly as $R^{-1/2}$ from $R=4$ to 64), and the solver converges in 12--15 Newton steps for $R\ge4$; a single vector without the floor does not converge.

\begin{table}
\caption{Maxent with sampled moments on DTW4096 (Kwon TB, $\Gamma$), no exact-moment substitution. Newton steps with the default floor $s_0=10^{-4}$ and with $s_0=0$ (``--'': not converged in 499 steps), and $L_1$ distance to the exact spectrum broadened by 0.05~eV (floor $10^{-4}$; the same for $s_0=0$ when it converges).}
\label{tab:rv}
\begin{ruledtabular}
\begin{tabular}{rrcccc}
 & & \multicolumn{2}{c}{Newton steps} & \\
$M$ & $R$ & $s_0=10^{-4}$ & $s_0=0$ & $L_1$ \\
\hline
100 & 1 & 29 & -- & 0.069 \\
100 & 4 & 12 & 12 & 0.046 \\
100 & 16 & 14 & 14 & 0.022 \\
100 & 64 & 15 & 15 & 0.014 \\
300 & 1 & 26 & -- & 0.126 \\
300 & 4 & 12 & 12 & 0.087 \\
300 & 16 & 14 & 14 & 0.038 \\
300 & 64 & 15 & 15 & 0.018 \\
\end{tabular}
\end{ruledtabular}
\end{table}

The sampling error of the band energy falls as $R^{-1/2}$ (Fig.~\ref{fig:acc}c) with $\approx270$~meV/atom per random vector for plain Rademacher vectors (193 at $R=2$, Table~\ref{tab:vec}); the default exact-moment substitution reduces this to $\approx16$~meV/atom per vector (11 at $R=2$).
Note that comparing a sharp Maxent curve with an exact spectrum broadened by a Gaussian penalizes Maxent for sharpness; the truncation test above avoids this by comparing integrated quantities and a narrow (0.05~eV) reference.

\begin{table*}[t]\caption{Truncation error on DTW4096 (Kwon, $\Gamma$) with \emph{exact} moments (no sampling noise): band-energy error and $L_1$ distance to the exact spectrum broadened by 0.05~eV. Maxent: exact-moment constraints; dashes: the solver did not reach tolerance $10^{-9}$ in 300 Newton steps, so no result is reported (no steps are listed either). Maxent $s_0$: every moment but $\mu_0$ given the uncertainty $s_0=10^{-4}$ (Sec.~\ref{sec:recon}); all converged.}\label{tab:acc}\begin{ruledtabular}\begin{tabular}{rrrrrrrrr}
$M$ & \multicolumn{3}{c}{$\Delta E_{\rm BS}$ (meV/atom)} & \multicolumn{3}{c}{pointwise $L_1$} & \multicolumn{2}{c}{Newton steps} \\ & KPM & Maxent & Maxent $s_0$ & KPM & Maxent & Maxent $s_0$ & Maxent & Maxent $s_0$ \\ \colrule
20 & $-198.23$ & $+22.52$ & $+22.52$ & 3.3493 & 0.1122 & 0.1122 & 9 & 9 \\
40 & $-18.24$ & $+1.56$ & $+1.56$ & 0.2862 & 0.0383 & 0.0383 & 11 & 11 \\
70 & $-0.99$ & $+0.12$ & $+0.10$ & 0.0715 & 0.0116 & 0.0114 & 34 & 17 \\
100 & $+0.263$ & -- & $+0.017$ & 0.0332 & -- & 0.0082 & -- & 22 \\
170 & $+0.297$ & -- & $-0.005$ & 0.0119 & -- & 0.0068 & -- & 23 \\
300 & $+0.138$ & -- & $-0.012$ & 0.0053 & -- & 0.0082 & -- & 24 \\
400 & $+0.085$ & -- & $-0.017$ & 0.0037 & -- & 0.0123 & -- & 22 \\
\end{tabular}\end{ruledtabular}\end{table*}

\subsection{Small systems: validity of exact-moment substitution}
\label{sec:small}

The test of Table~\ref{tab:vec} is one system with $N=16\,384$. I repeated the comparison where the argument for substitution could fail, with exact references from dense diagonalization: a periodic chain and a dimerized (gapped) chain with $N=32$--4096, the Kwon tight-binding Hamiltonian of cubic diamond silicon cells $2^3$--$5^3$ ($N=256$, 864, 2048, 4000 orbitals, $\Gamma$) and the three-dimensional Anderson model ($W=3$, $N=512$, 1000, 4096). The moments of random vectors were computed in the eigenbasis, which gives the same numbers as the recursion. Errors of the band energy $E_{\rm BS}$ (half filling) and of the cumulative distribution are measured against the same reconstruction fed with the exact moment list, so that truncation is common to all estimators.

\begin{figure}
\includegraphics[width=\columnwidth]{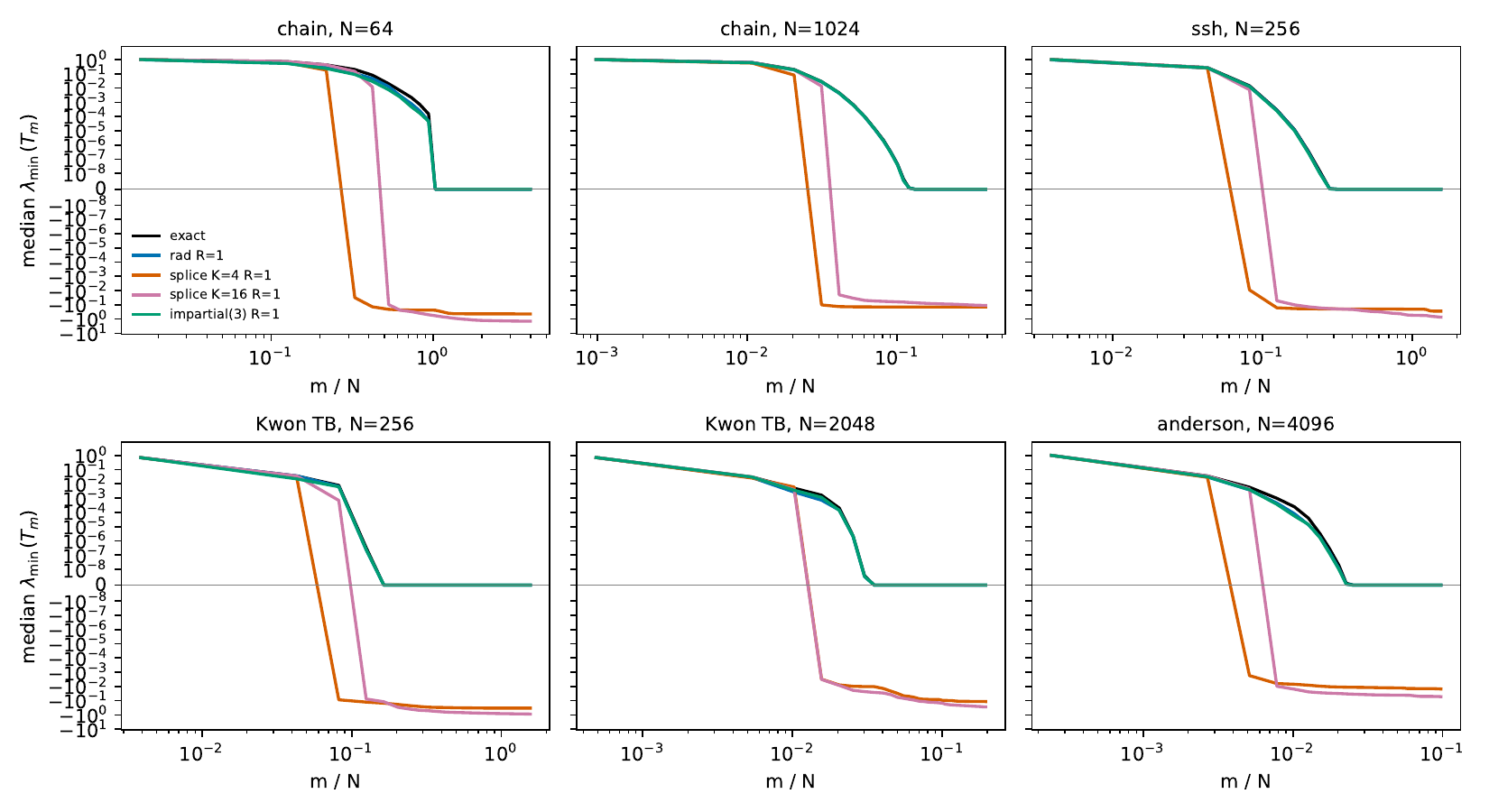}
\caption{Smallest eigenvalue of the Toeplitz matrix $T_m$ (median over seeds) against $m/N$ (Kwon TB: the tight-binding Hamiltonian of Kwon \emph{et al.}): exact moments and plain or impartial vectors (all curves on top of each other, never negative) and spliced lists $K=4$, 16 (negative beyond $m\approx12$--$40$).}\label{fig:cone}
\end{figure}

\emph{Consistency.} A list $\mu_0,\dots,\mu_m$ is the moment list of a positive measure on $[-1,1]$ if and only if the Toeplitz matrix $T_{jk}=\mu_{|j-k|}$ is positive semidefinite. The exact lists satisfy $\lambda_{\min}\ge-6\times10^{-14}$. Lists from one random vector or an average of 16 vectors, and impartial vectors, never violate it (all seeds, all systems). The spliced lists ($K=2$--16, one or 16 vectors) violate it in 80--100\% of the seeds, with $\lambda_{\min}$ down to $-0.02$ to $-0.7$, at an order $m^*\approx12$--$40$ that does not grow with $N$ (Fig.~\ref{fig:cone}); the empty arcs of the spectral measure (band-edge margin, gaps) make $\lambda_{\min}$ of the exact list small at low order, and the $N^{-1/2}$ noise of the sampled tail exceeds it.

\begin{figure*}
\includegraphics[width=\textwidth]{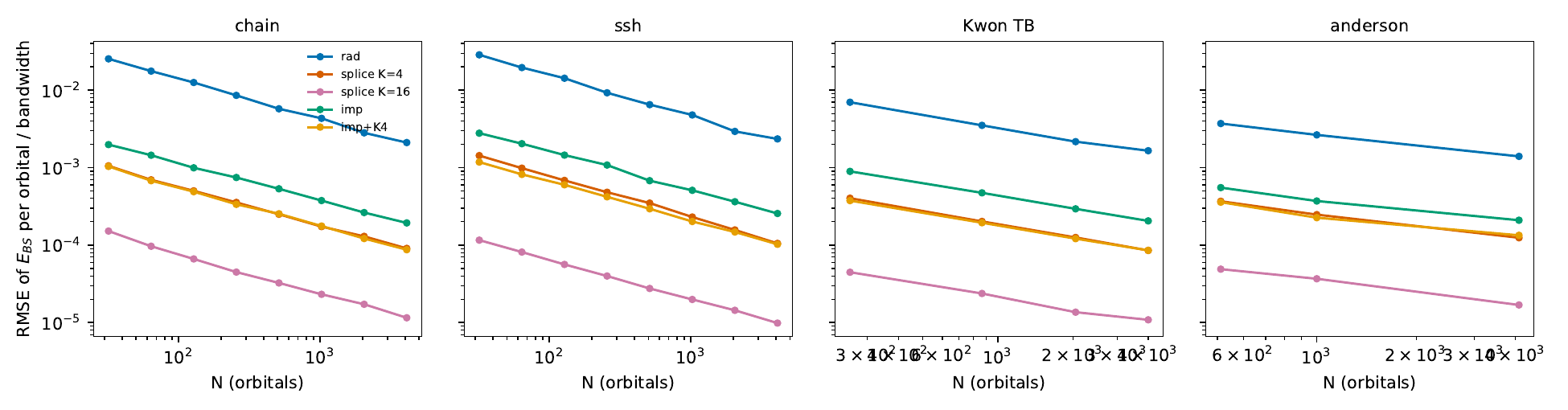}
\caption{Root-mean-square error of the KPM band energy per orbital, divided by the bandwidth, against $N$ ($M=100$, $R=4$; 500 seeds for $N\le1024$, 200 above), for plain Rademacher vectors, exact-moment substitution ($K=4$, 16), impartial vectors and impartial vectors with $K=4$.}\label{fig:rmse}
\end{figure*}

\emph{KPM.} At equal cost the root-mean-square error of $E_{\rm BS}$ ($M=100$, $R=4$) is $0.04$--$0.10$ ($K=4$) and $0.004$--$0.014$ ($K=16$) times that of plain Rademacher vectors, for every system and $N$ from 32 to 4000; impartial vectors give $0.08$--$0.15$ (Fig.~\ref{fig:rmse}). The estimates of $E_{\rm BS}$ have a bias that is statistically significant with 500 seeds, for plain vectors as well as spliced ones, but much smaller than the plain-vector error. Its origin is the Fermi level: with $E_F$ held at its exact value, $E_{\rm BS}$ is a linear functional of the moments and is unbiased to within two standard errors for plain vectors and for $K=4$, 16 (2000 seeds). The impartial vectors are unbiased at the level of one standard error for $N\gtrsim500$ and show an rms $z$-score of $1.3$--$1.6$ for $N\lesssim100$. Constraining more moments (script \texttt{zscore\_constrained\_impartial.py}, 1000--1500 vectors, rms $z$ of the unconstrained moments $n$ above the constrained ones and below 64) gives $1.0$--$1.5$ for 4, 8 and 16 constrained moments ($1.5$ for the Kwon TB with 864 orbitals, $1.3$--$1.4$ for the Anderson model with 1000 sites, about $1.0$ for the Kwon TB with 2048 orbitals and for a chain with 512 sites), against $0.8$--$1.0$ for the three constrained moments of Drabold and Sankey: a bias of at most about one standard error of the mean of the vectors at these sizes.

\emph{Maxent.} With sampled uncertainties from $R=16$ vectors, substitution with $K=4$ lowers the error of $E_{\rm BS}$ by 6--40 times relative to plain vectors and Maxent converges in 90--100\% of the runs ($K=16$ needs 20--66 Newton steps, against 11--17 for plain vectors). With a single vector the uncertainty is only the floor, the inconsistent list has no solution, and Maxent with $K=4$ or 16 does not converge (0--20\% of the runs reach the tolerance in 500 steps) and returns a band energy and a cumulative distribution that are far worse than with plain vectors, at every size tested (chain, dimerized chain, Kwon, Anderson). Plain and impartial vectors converge with $R=1$. These $R=1$ and $R=16$ runs used $M=100$, the chain and the dimerized chain with $N=64$ and 512, Kwon with $N=256$ and 864 and Anderson with $N=512$, with 20 seeds (5 for the non-converging $R=1$ splices); the runs without the uncertainty floor and with $M>N$ were not repeated. For $R=2$, 3, 4 and 8 (script \texttt{e2\_maxent\_R.py}; $M=100$; chain and dimerized chain with $N=512$, Kwon with $N=256$ (60 seeds) and 864, Anderson with $N=512$, 40 seeds each unless stated) Maxent with $K=4$ converged in 90--100\% of the runs (typically 95--98\%, with no trend in $R$), in 9--19 Newton steps (median), and lowered the rms error of $E_{\rm BS}$ by factors of 8--40 against plain vectors with the same $R$; plain vectors converged in 95--100\% of the runs. The failure is therefore specific to $R=1$. These runs were limited to $N\le864$.

\emph{Resolution at the level spacing.} For $N=256$ and 1024, KPM with $M$ up to $4N$ stays stable and its error is set by the stochastic noise, not by $M$. Maxent from the exact moment list, with the default floor, needs 42--129 Newton steps for $M\le N$ and did not converge at $M=N$ for the Kwon system with $N=256$. The package warns when $M\ge N$ and suggests dense diagonalization, which is cheaper and exact for such small problems.

\subsection{Complex Hermitian problems}
\label{sec:val_cplx}

A Bloch twist $\mathbf k$ makes the Hamiltonian of a periodic supercell complex Hermitian, $H_{ij}=h(\mathbf d_{ij})e^{i\mathbf k\cdot\mathbf d_{ij}}$, and its spectrum is the union of the primitive-cell bands at the $\mathbf k$ points folded onto it.
For a 64-atom Kwon supercell and a 216-atom DFTB supercell (the generalized problem), the spectrum of the twisted supercell equals this union to $10^{-9}$ and $10^{-8}$~eV, respectively, and $H$ is Hermitian to $10^{-12}$; the same builder is used in chunked form for $10^6$ atoms and agrees with the unchunked one exactly.
The matrix-free operator applies $H$ as a stencil: the product with a vector is a sum of 128 shifted $4\times4$ blocks (one for each neighbour pair of the cell), so that $H$ is never stored. For a $16^3$-cell supercell (131\,072 orbitals) at a generic twist the stencil has a Hermiticity defect of $1.8\times10^{-17}$ and its KPM and Maxent densities ($M=170$, $R=6$) agree with the exact $\mathbf k$-sum with $L_1$ distances $0.036$ and $0.047$ against the 0.1~eV-broadened reference.
For DTW4096 with the DFTB model the density at $\mathbf k=(\mathbf b_1+\mathbf b_2+\mathbf b_3)/4$ ($\mathbf b_i$ the reciprocal-lattice vectors of the cell; a quarter of each, so $\mathbf k$ is not a reciprocal-lattice vector and the matrices are not gauge-equivalent to those at $\Gamma$), a genuinely complex generalized problem, looks like the $\Gamma$ one ($L_1$ to the $\Gamma$ spectrum $0.021$ against $0.024$ for $\Gamma$ itself, Table~\ref{tab:dd}); a dense complex generalized diagonalization of 16\,384 orbitals does not fit in this machine, so there is no exact reference at that $\mathbf k$ point (at a reciprocal-lattice vector $\mathbf G$ the twisted matrices are unitarily equivalent to those at $\Gamma$, which would give an exact reference; I did not use such a point).

\subsection{Non-orthogonal problems: DFTB and SIESTA}
\label{sec:val_dftb}

\paragraph*{Comparison with DFTB+.}
DFTB+ \cite{dftbplus} can write the real-space Hamiltonian $H$ (the zeroth-order, non-self-consistent one) and $S$ (\texttt{WriteRealHS}).
For a randomly displaced 64-atom silicon cell of 10.9~\AA\ (where pairs interact through two periodic images) the matrices built by \sdos\ from the same \texttt{.skf} files agree with DFTB+ to $7.9\times10^{-7}$~eV in $H$ (largest element 10.8~eV) and $4.3\times10^{-8}$ in $S$, and the generalized eigenvalues to $3\times10^{-6}$~eV.
A distorted Si--H cluster (which tests the heteronuclear sign convention and the use of the reversed-pair file) agrees to $7.3\times10^{-7}$~eV and $5.5\times10^{-8}$.
The residual is the difference between my cubic spline and DFTB+'s polynomial interpolation; in the interior of the table the two agree to $10^{-10}$.

\paragraph*{DTW4096.}
Table~\ref{tab:dd} gives the DFTB problem of DTW4096: 16\,384 orbitals with 179 nonzeros per row, dense generalized spectrum computed exactly (Cholesky reduction and a dense eigensolver, 210~s, about three dense matrices of 2.1~GB each), gap $0.0095$~eV.
The three ways of applying $S^{-1}$ give the same moments (identical band energy). The error against the dense generalized spectrum is a sampling error, not a bias. For $R=6$ and four seeds it is $-70$, $-122$, $+69$ and $+153$~meV/atom (rms 110), for $R=24$ and two seeds $-3$ and $+113$ (rms 80), and for $R=96$ (one run) $+2$~meV/atom (script \texttt{dtw\_dftb\_Rscan.py}). This is consistent with the $R^{-1/2}$ law (expected rms 55 and 27~meV/atom from the $R=6$ value), within the small number of runs. To test unbiasedness directly I ran eight seeds at $R=96$ with the Chebyshev $B^{-1}$ mode and $M=100$ (about 9~min each; script \texttt{dtw\_dftb\_unbiased.py}), and compared every stochastic moment with the dense generalized moment: the $z$-scores (stochastic minus exact, over the sample standard error) have rms $0.82$--$1.22$ per seed (pooled $1.00$ over 792 moments) and mean $0.006$ (the standard error would be $1/\sqrt{792}=0.036$ for independent moments; the eight per-seed means scatter by only $0.011$, so the moments are not independent and I quote no further error), and $\Delta E_{\rm BS}$ (the KPM band energy at the same $M$, so truncation cancels) is $-0.3\pm7.7$~meV/atom (seed spread 22, range $-42$ to $+25$), consistent with zero bias and with the $R^{-1/2}$ law. The standard error estimated from the vectors of a single run is itself uncertain at small $R$ (35 to 94~meV/atom at $R=6$), so Table~\ref{tab:dd} should be read with the seed spread, not its $\pm$. For the intended use this means that $R=6$ vectors give the shape of the density of states but a band energy to only $\approx\pm0.1$~eV/atom, and the energy differences between structures that matter in a workflow need either many more vectors (the cost is linear in $R$) or an independent estimate of the error; the variance reductions of Table~\ref{tab:vec} are not available here, because exact substitution requires a diagonal $B$ and impartial vectors are refused for a generalized problem. Using the same random vectors for two similar structures should cancel part of the noise; I have not tested it. IC(0) halves the number of CG iterations ($11$ against $24$ per solve) but is slower in wall-clock time because of its setup.

\paragraph*{SIESTA.}
The SIESTA Harris/single-zeta matrices of the fullerenes C$_{240}$ and C$_{540}$ (960 and 2160 orbitals), generated with SIESTA 5.4.2 \cite{siesta2020} and read through \texttt{sisl}, give HOMO--LUMO gaps from the inertia bisection that agree with the dense \texttt{eigh}$(H,S)$ gaps to $2\times10^{-10}$ and $4\times10^{-11}$~eV (1.2995 and 0.9220~eV).
For the series C$_{60n^2}$ up to C$_{2160}$ the PCG needs $59$ iterations per solve at every size and the cost is linear in $N$; SIESTA itself ran out of memory at C$_{3840}$ (12~GB) because of its $N^2$ setup, which is a reason to build $H,S$ by the sparse route above.

\begin{table*}[t]\caption{DFTB electronic problem of DTW4096 (16\,384 orbitals, 179 nonzeros per row): KPM, $M=170$, $R=6$. At $\Gamma$ the three ways of applying $S^{-1}$ give identical band energies (the same moments); the error is against the dense generalized spectrum and is a sampling error (this run, seed 11: $-70$~meV/atom, with $\pm35$ the standard error estimated from its six vectors; other seeds give $-122$, $+69$ and $+153$~meV/atom with estimated errors of 84--94, so that estimate is itself uncertain and the spread over seeds, about 110~meV/atom at $R=6$, is the reliable figure; see text), and $L_1$ is the distance to the exact spectrum broadened by 0.1~eV. The dense complex generalized problem does not fit in memory here, so the $\mathbf k$ point is compared with the $\Gamma$ spectrum of the same cell.}\label{tab:dd}\begin{ruledtabular}\begin{tabular}{lrrrrrrr}
$S^{-1}$ by & \multicolumn{4}{c}{$\Gamma$ (real)} & \multicolumn{3}{c}{$\mathbf k$ (complex Hermitian)} \\ & time (s) & iter./solve & $\Delta E_{\rm BS}$ (meV/atom) & $L_1$ to exact & time (s) & iter./solve & $L_1$ to $\Gamma$ exact \\ \colrule
PCG, block-Jacobi & 54 & 23.6 & $-70.0\pm35$ & 0.024 & 125 & 23.6 & 0.021 \\
PCG, IC(0) & 79 & 11.0 & $-70.0\pm35$ & 0.024 & 155 & 11.0 & 0.021 \\
Chebyshev polynomial & 64 & 26.0 & $-70.0\pm35$ & 0.024 & 135 & 26.0 & 0.021 \\
\end{tabular}\end{ruledtabular}\end{table*}

\subsection{Vibrations}
\label{sec:val_vib}

\begin{figure*}
\includegraphics[width=\textwidth]{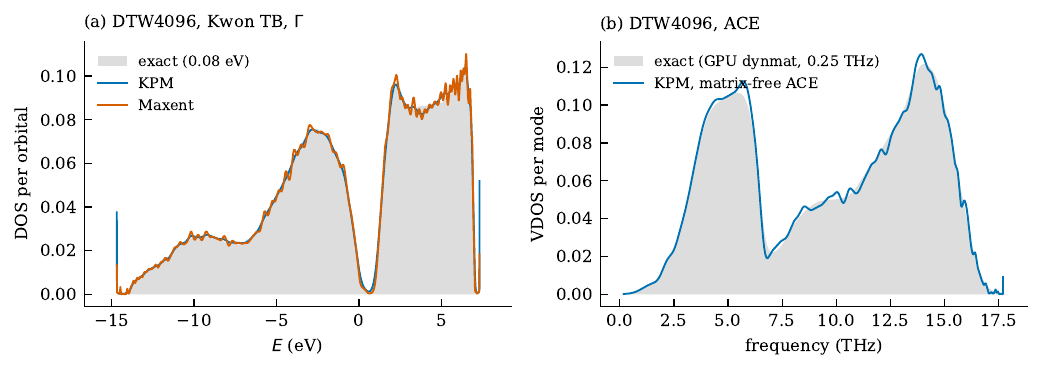}
\caption{DTW4096 (a 4096-atom model of a-Si) at $\Gamma$. (a) Electronic density of states (Kwon TB), KPM and Maxent ($M=170$, $R=12$) against the exact spectrum. (b) Vibrational density of states with the ACE potential: matrix-free KPM ($M=170$, $R=4$, two force calls per product) against the exact dynamical matrix of LAMMPS (grey).}
\label{fig:dtw}
\end{figure*}

\paragraph*{The operator.}
For a 512-atom silicon structure with the ACE potential \cite{ace,pace} for the silicon--hydrogen system of Rosset and Deringer \cite{rosset2026} the matrix-free products agree with the exact dynamical matrix written by LAMMPS' \texttt{dynamical\_matrix} command to a relative $2.4\times10^{-5}$ (the expected $O(\delta^2)$ error of the finite difference) with the same units, using two force calls per product.
\paragraph*{DTW4096.}
For the 4096-atom cell (3 rigid translations removed exactly) the matrix-free results agree with the exact dense spectra of both potentials (Table~\ref{tab:ph}): for Stillinger--Weber \cite{sw} (dense matrix built column by column through the same force interface) the zero-point energy is $0.5\%$ and $C_v(300\,\mathrm{K})$ $0.3\%$ off; for ACE, whose exact matrix I obtained with LAMMPS (Kokkos on the GPU, \texttt{pair\_style pace product}) in about ten minutes, the ZPE is $-0.66\%$ and $C_v(300\,\mathrm{K})$ $+0.27\%$ off, and the exact matrix has no negative eigenvalues and exactly three zero modes.
The largest error is the heat capacity at 100~K ($+1.4\%$), where the result depends on the lowest frequencies, which have the poorest resolution in $\lambda=\omega^2$.
Fig.~\ref{fig:dtw}b shows the density: the acoustic band peaks near 5~THz ($\approx21$~meV) and the optic band near 14~THz ($\approx58$~meV).
KPM and Maxent agree to four digits for ACE and to three for Stillinger--Weber.

\begin{table*}[t]\caption{Harmonic thermodynamics of DTW4096 (4096 atoms, relaxed with each potential): matrix-free sdos ($M=170$, $R=4$, two force calls per product, three rigid translations removed exactly) against the exact dynamical matrix. $C_v$ in $k_B$/atom.}\label{tab:ph}\begin{ruledtabular}\begin{tabular}{llrrrr}
potential & method & ZPE (meV/atom) & $C_v$(100 K) & $C_v$(300 K) & $C_v$(600 K) \\ \colrule
SW & exact (dense) & 66.935 & 0.8558 & 2.2935 & 2.7863 \\
SW & sdos KPM & 66.573 & 0.8637 & 2.3002 & 2.7890 \\
SW & sdos Maxent & 66.546 & 0.8647 & 2.3007 & 2.7892 \\
\colrule
ACE & exact (GPU \texttt{dynamical\_matrix}) & 60.104 & 0.9369 & 2.4127 & 2.8294 \\
ACE & sdos KPM & 59.709 & 0.9499 & 2.4192 & 2.8319 \\
ACE & sdos Maxent & 59.709 & 0.9499 & 2.4193 & 2.8319 \\
\end{tabular}\end{ruledtabular}\end{table*}

\subsection{Windows and gaps}

On DTW4096 the shift-invert window around the gap returns the same three levels as the dense spectrum to $7\times10^{-14}$~eV.
The exact fullerene gaps from the inertia bisection are in Table~\ref{tab:fu}.

\section{Applications}
\label{sec:apps}

\subsection{A 2\,097\,152-atom crystal of silicon}
\label{sec:app_si}

\begin{figure*}
\includegraphics[width=\textwidth]{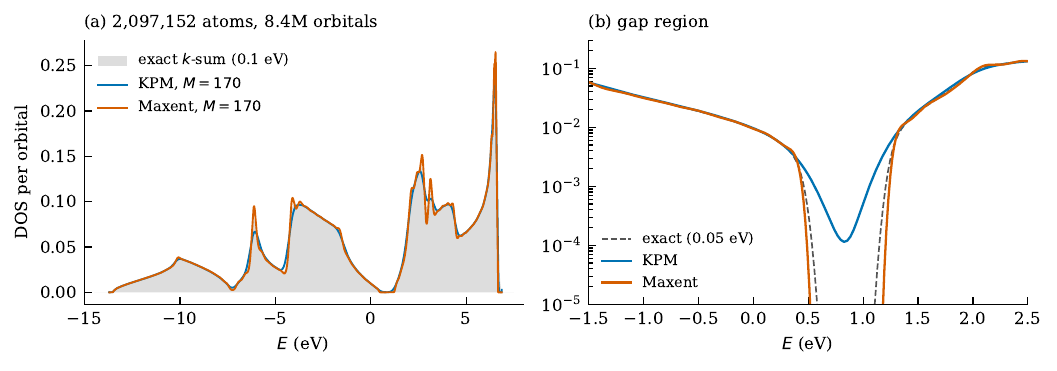}
\caption{Kwon tight-binding DOS of diamond silicon with $64^3$ cubic cells (2\,097\,152 atoms, 8\,388\,608 orbitals), $M=170$ moments, $R=4$ vectors, matrix-free; grey: exact $\mathbf k$-sum. (a) Whole band. (b) Gap region on a logarithmic scale.}
\label{fig:si2m}
\end{figure*}

Extended Data Fig.~9 of Deringer \emph{et al.}\ \cite{deringer} shows the tight-binding DOS of a cubic diamond supercell of ``2\,097\,172'' atoms (34.7584~nm on a side) computed with the Kwon tight-binding Hamiltonian \cite{kwon}, 170 moments and Maxent.
A cell of 34.7584~nm is exactly $64^3$ cubic cells of 0.5431~nm, \ie\ $2\,097\,152$ atoms (Ref.~\cite{deringer}, of which I am a coauthor, prints 2\,097\,172), and the matrix has 8\,388\,608 orbitals.
Because the cell is periodic and translation invariant, the Hamiltonian is applied as a stencil (128 shifted $4\times4$ blocks per vector) and never stored.
With $M=170$ and $R=4$ vectors the KPM density takes 603~s on 4 threads ($358$~s for the bounds, $246$~s for the moments), with a peak of 1.8~GB; the Maxent density from the same moments takes the same time (the fit adds less than 1~s); Fig.~\ref{fig:si2m}.
The exact reference is the 8-atom cell bands on the $64^3$ mesh (exactly the $\Gamma$ spectrum of the supercell), a histogram with exactly 8\,388\,608 orbitals.
Both methods reproduce the exact spectrum. The valence peaks are at $-6$ and $-4$~eV, the conduction features at $+2$ to $+3$~eV and the sharp upper edge at 6.7~eV, as in the published figure. The band runs from $-13.5$ to $6.65$~eV.
The methods differ most in the gap (Fig.~\ref{fig:si2m}b). The exact band edges on this mesh are $0.46$ and $1.24$~eV, and Maxent follows them: its density is below $10^{-3}$ between $0.44$ and $1.24$~eV and below $10^{-4}$ between $0.49$ and $1.20$~eV. KPM at 170 moments (resolution $\pi a/M=0.19$~eV, $a=10.28$~eV; FWHM $0.45$~eV) fills the gap down to $10^{-4}$ (minimum $1.1\times10^{-4}$ at $0.82$~eV), a Jackson-broadened floor. For anything that depends on the gap this is the advantage of Maxent for integrated quantities, noted in Sec.~\ref{sec:val_acc}.
The same calculation takes 60~s at $32^3$ cells (262\,144 atoms) and 264~s at $48^3$ cells (884\,736 atoms). The time grows with the exponent $1.12$ against $1.01$ for the explicit matrices of Sec.~\ref{sec:scaling}: $1.09$ for the moments and $1.13$ for the bounds, whose ARPACK iteration needs more products as the system grows ($162$, $222$, $372$, $392$ and $522$ at $8^3$, $16^3$, $24^3$, $32^3$ and $48^3$ cells, against 340 for the moments). For a crystal the bounds can be given analytically.

\subsection{Fullerenes and graphene}
\label{sec:app_fu}

\begin{figure*}
\includegraphics[width=\textwidth]{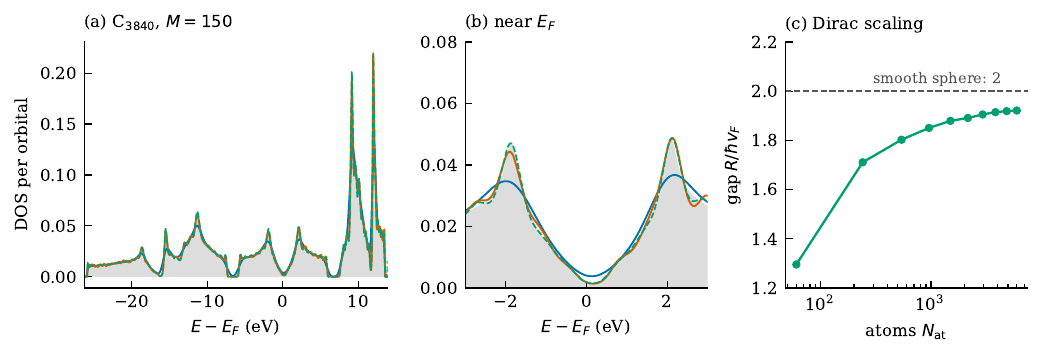}
\caption{Fullerene C$_{3840}$ with the Xu Hamiltonian on an LCBOP-relaxed geometry. (a,b) DOS from $M=150$ moments (blue: KPM; red: Maxent) against the exact eigenvalues (grey, 0.1~eV Gaussian), full band and near $E_F$; green dashed: Maxent with the floor $s_0=10^{-4}$ from the exact moments, $M=150$. (c) $\mathrm{gap}\,R_{\rm eq}/\hbar v_F$ against the number of atoms; dashed: the smooth-sphere value 2.}
\label{fig:fu}
\end{figure*}

Drabold, Ordej\'on, Dong and Martin \cite{ssc96} followed the icosahedral fullerenes C$_{60n^2}$ from C$_{60}$ to C$_{3840}$ with moment-Maxent densities of states (``of order 150 moments'', no artificial broadening), a shifted-Lanczos HOMO--LUMO gap and local densities of states, and observed that the DOS of C$_{2160}$ and C$_{3840}$ approaches graphene; the geometries were relaxed with an order-$N$ tight-binding scheme \cite{itoh}.
I repeat the calculation with the Xu--Wang--Chan--Ho Hamiltonian \cite{xu} on geometries relaxed with the LCBOP potential \cite{lcbop} (which reproduces the strain energies and radii of Ref.~\cite{itoh} to within a few percent, while REBO and AIREBO overestimate the strain by 35--70\%), generated with a Caspar--Klug (Goldberg) construction for any $(m,k)$.
Fig.~\ref{fig:fu}a shows C$_{3840}$ from $M=150$ moments. Maxent follows the exact spectrum (a 0.1~eV Gaussian-broadened histogram of the 15\,360 exact eigenvalues, computed in 124~s), including the van Hove peaks near $+9$ and $+12$~eV above $E_F$. KPM at 150 moments (resolution $\pi a/M\approx0.42$~eV, $a=19.9$~eV) is visibly broader. This is the regime of the 1995 calculation, and the sharp graphene-like features it showed are reproduced.
The Maxent curve in Fig.~\ref{fig:fu}a,b uses the default uncertainty floor $s_0=10^{-4}$; with 12 random vectors the sampled errors are larger and the floor is inactive. To see what $s_0$ does here I also fitted exact moments, computed from the exact spectrum, at $M=150$ (green dashed line). The $L_1$ distance to the exact spectrum, broadened by 0.1~eV, is $0.064$ with the floor and $0.065$ without it (the plain solver converges here, in 403 steps instead of 27), against $0.070$ for Maxent with the 12 random vectors and $0.099$ for KPM. At this $M$ the floor is therefore free, and the stochastic error costs Maxent little.
The exact HOMO--LUMO gap of C$_{3840}$ in the Xu model is $0.284$~eV against $0.10$~eV in their Fig.~3 (and $1.537$ against $1.62$~eV for C$_{60}$). The difference may come from the different relaxations of the geometry (the cages of Ref.~\cite{ssc96} were relaxed with the order-$N$ scheme of Ref.~\cite{itoh}, mine with LCBOP), but the comparison with a sphere below argues against this, since a much larger change of geometry changes the gap by less than 1\% at C$_{3840}$; I did not resolve it, and the shifted-Lanczos gap of 1995 may not have been converged.

\begin{figure*}
\includegraphics[width=0.85\textwidth]{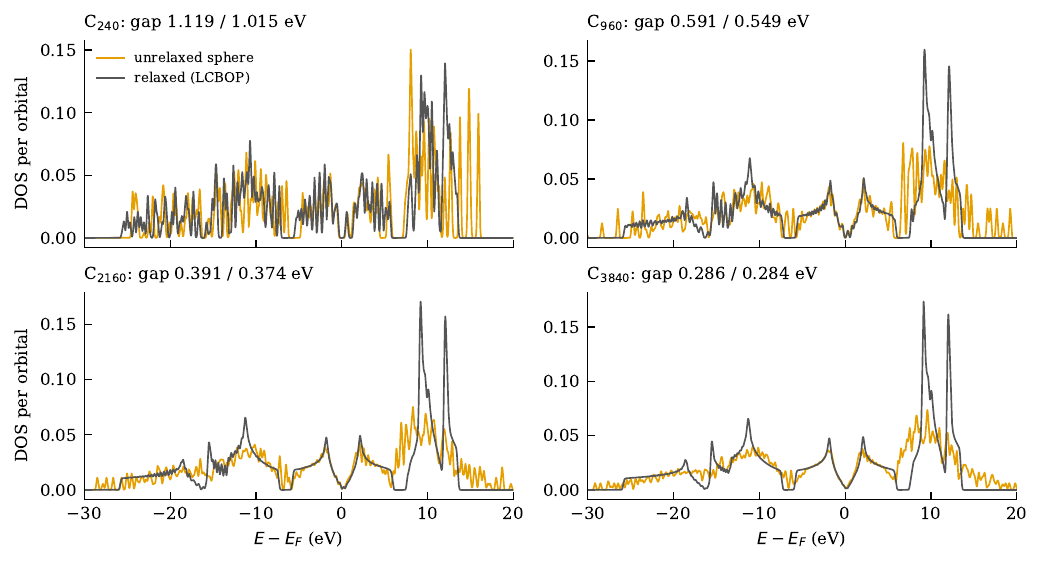}
\caption{Exact spectra of C$_{240}$, C$_{960}$, C$_{2160}$ and C$_{3840}$ with the Xu Hamiltonian (0.1~eV Gaussian broadening): LCBOP-relaxed geometry (grey) and the unrelaxed spherical Caspar--Klug geometry (orange). The HOMO--LUMO gaps (sphere / relaxed) are given in the panel titles.}
\label{fig:fusph}
\end{figure*}

To see what the relaxation does, the same Hamiltonian was employed on the unrelaxed spherical geometry: the Caspar--Klug construction projected onto a sphere of the graphene-equivalent radius ($28.29$~\AA). All atoms are then at the same distance from the centre, and the bond lengths scatter between $1.13$ and $1.56$~\AA\ (rms $0.10$~\AA). The LCBOP geometry has bonds between $1.41$ and $1.45$~\AA\ (rms $0.003$~\AA), but its distances from the centre range from $26.9$ to $31.6$~\AA\ (rms $1.1$~\AA): the relaxed cage is faceted. The two exact spectra of C$_{3840}$ (Fig.~\ref{fig:fusph}) have almost the same gap ($0.286$ and $0.284$~eV), with $93$ and $72$ levels within $0.6$~eV of $E_F$, but otherwise differ strongly. The same comparison for smaller cages (Fig.~\ref{fig:fusph}) shows the gap of the sphere larger than the relaxed one by $10\%$ at C$_{240}$ ($1.119$ and $1.015$~eV), $8\%$ at C$_{960}$ ($0.591$ and $0.549$~eV) and $4\%$ at C$_{2160}$ ($0.391$ and $0.374$~eV), and the van Hove peaks of the relaxed cages sharper at every size (those of the sphere are smeared); the band of the sphere is also wider ($-42$ to $+25$~eV against $-22$ to $+17$~eV) because of its short bonds. The sharp graphene-like features of the 1995 spectrum thus require the flat regions of a faceted cage and do not occur on a uniformly curved one. Part of the smearing on the sphere is bond-length disorder, which is the strain that the relaxation removes, so this comparison does not separate curvature from strain. The gap, a property of the lowest levels near the Dirac point, is insensitive to both. Maxent from 150 moments (stochastic, $R=12$) follows the exact sphere spectrum with an $L_1$ distance of $0.118$ against $0.155$ for KPM.

The exact gaps (Table~\ref{tab:fu}) follow $\mathrm{gap}\propto N_{\rm at}^{-0.47}$ (fit over C$_{240}$--C$_{6000}$; fits starting at C$_{60}$, C$_{960}$ and C$_{2160}$ give $-0.43$, $-0.48$ and $-0.48$), which is the scaling of a confined massless Dirac fermion, $\hbar v_F/R$ with $R\propto N_{\rm at}^{1/2}$.
The graphene bands of the same Hamiltonian have a true Dirac point (gap at K $<10^{-14}$~eV) with $\hbar v_F=4.20$~eV\,\AA\ ($6.4\times10^5$~m/s). The product $\mathrm{gap}\cdot R_{\rm eq}/\hbar v_F$ rises with $N_{\rm at}$ to $1.92$ at C$_{6000}$ (Fig.~\ref{fig:fu}c), close to the value $2$ of a smooth sphere without pentagons, whose lowest levels are $\pm\hbar v/R$. The twelve pentagons carry a fictitious gauge flux that modifies the continuum spectrum \cite{gonzalez1992,gonzalez1993}. The $4\%$ is therefore read as agreement of the scaling, not as a test.
The weight of the states within $0.6$~eV of $E_F$ on the 12 pentagon neighbourhoods (3\% of the atoms), the 30 edge strips and the 20 facet interiors (shift-invert on C$_{960}$, C$_{2160}$, C$_{3840}$; weights summed over each Ih multiplet) shows that the pentagons matter much more than the edges: for C$_{3840}$ the pentagon enhancement ranges between $0.26$ and $3.4$ across multiplets, the edge and interior enhancements between $0.75$ and $1.3$; the states stay extended (IPR $\sim1/N$).
The class widths (2.5 and 3~\AA) are my choice and the edge distance uses the mean radius of a faceted cage, so the ranking of the multiplets is more reliable than the numbers.

\begin{figure*}
\includegraphics[width=0.85\textwidth]{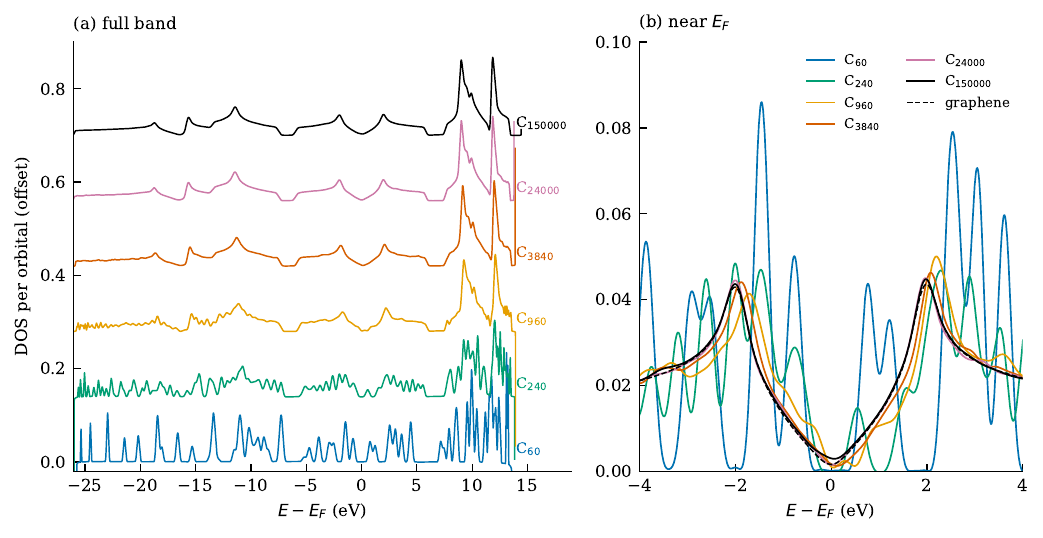}
\caption{Density of states of the fullerenes C$_{60}$ to C$_{150000}$ (LCBOP-relaxed geometries, Xu Hamiltonian, KPM with $M=1500$ moments, resolution $\pi a/M\approx0.042$~eV) against the DOS of graphene from the same Hamiltonian (dashed). (a) Full band, curves offset vertically. (b) Near $E_F$.}
\label{fig:fuseries}
\end{figure*}

Figure~\ref{fig:fuseries} shows how the density of states evolves with the size of the fullerene, from C$_{60}$ to C$_{150000}$: the molecular levels of C$_{60}$ merge into bands, the van Hove peaks sharpen, and near $E_F$ the DOS of C$_{150000}$ cannot be distinguished from that of graphene on the scale of the figure.
For the cost, the 600\,000-orbital problem C$_{150000}$ takes 372~s and 1.35~GB for 1500 moments and two vectors (timing on an unrelaxed ideal-sphere geometry, valid as a cost but not as a spectrum; the DOS of Fig.~\ref{fig:fuseries} is for the relaxed geometry and took 471~s), with 37 nonzeros per row and the exact-gap bisection capped at 25\,000 orbitals because of fill.

\begin{table}[t]\caption{Exact HOMO--LUMO gaps (inertia bisection, Xu Hamiltonian, LCBOP-relaxed icosahedral fullerenes) and the confined-Dirac-cone scaling; $\hbar v_F=4.196$~eV\,\AA\ from the graphene bands of the same Hamiltonian, $R_{\rm eq}=(N_{\rm at} A_0/4\pi)^{1/2}$ with $A_0$ the graphene area per atom.}\label{tab:fu}\begin{ruledtabular}\begin{tabular}{rrrr}
C$_{N_{\rm at}}$ & gap (eV) & $R_{\rm eq}$ (\AA) & gap$\,R_{\rm eq}/\hbar v_F$ \\ \colrule
60 & 1.537 & 3.54 & 1.295 \\
240 & 1.015 & 7.07 & 1.711 \\
540 & 0.713 & 10.61 & 1.803 \\
960 & 0.549 & 14.15 & 1.851 \\
1500 & 0.446 & 17.68 & 1.879 \\
2160 & 0.374 & 21.22 & 1.891 \\
2940 & 0.323 & 24.76 & 1.906 \\
3840 & 0.284 & 28.29 & 1.915 \\
4860 & 0.253 & 31.83 & 1.919 \\
6000 & 0.228 & 35.36 & 1.922 \\
\end{tabular}\end{ruledtabular}\end{table}

\subsection{Amorphous silicon with DFTB and ACE}
\label{sec:app_dtw}

Figure~\ref{fig:dtw} and Tables~\ref{tab:dd}, \ref{tab:ph} summarize DTW4096 as a stress test of both branches: the electronic DOS from the Kwon Hamiltonian (exact gap $0.0149$~eV at $\Gamma$) and the DFTB problem (generalized, real and complex, gap $0.0095$~eV), and the vibrational DOS with the ACE potential \cite{rosset2026} that relaxed the structure.
The small gaps are properties of this $\Gamma$-point cell, not of the method; the zero-point energy is $60.1$~meV/atom and $C_v(300\,\mathrm{K})=2.41\,k_B$/atom.
The tiling of DTW4096 into $2^3$ cells with the SW potential reproduces the single-cell ZPE per atom to $2\times10^{-5}$ (67.0505 and 67.0516 meV/atom for 4096 and 32\,768 atoms; matrix-free KPM with $M=100$, $R=2$ on the same SW-relaxed structure as Table~\ref{tab:ph}). The same structure at $M=170$, $R=4$ gives 66.573~meV/atom (Table~\ref{tab:ph}) against 66.935 exact, so the dependence on $M$ and $R$ (the $M=100$ run used $R=2$, the $M=170$ run $R=4$) is larger than the difference between the two cells.

\begin{figure*}
\includegraphics[width=\textwidth]{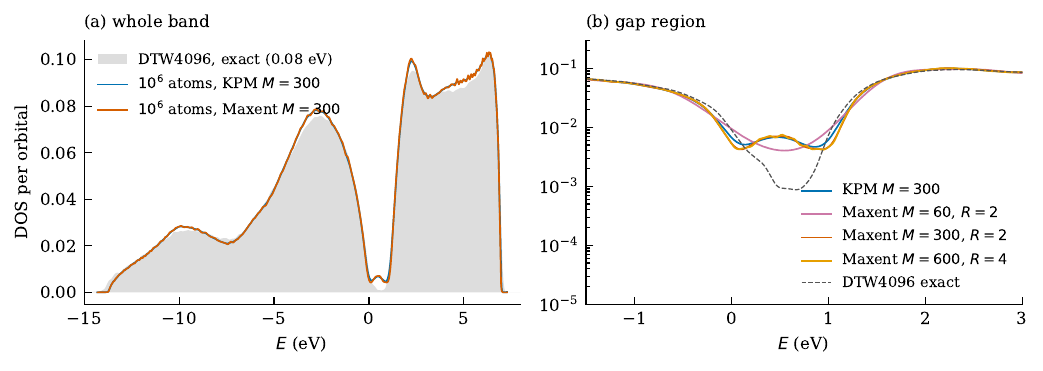}
\caption{\label{fig:morrow}DOS of the one-million-atom a-Si model of Morrow \emph{et al.}\ \cite{morrow2024,morrowdata} with the Kwon tight-binding Hamiltonian ($\Gamma$ point, 4 million orbitals). (a) KPM and Maxent, $M=300$, $R=2$, against the exact 4096-atom DTW spectrum (0.08~eV broadening). (b) Gap region on a log scale; Maxent from $M=60$ to 600 ($R=2$ and 4) and DTW4096.}
\end{figure*}

Large amorphous models have been treated with this kind of method before: the Supplementary Information of Morrow \emph{et al.}\ \cite{morrow2024} shows the electronic density of states of a one-million-atom amorphous-silicon model (their Supplementary Fig.~3a in the arXiv version, arXiv:2308.16868), computed by the maximum-entropy method of Ref.~\cite{drabold1993} and compared for increasing numbers of moments, and a gap similar to that of a 512-atom model treated with a hybrid functional.
I repeated the calculation for the same structure, using the one-million-atom main-text model deposited by those authors \cite{morrowdata} (cubic cell of 273.6~\AA, periodic, $\Gamma$ point; file \texttt{aSi\_1Matom\_maintext.xyz})\footnote{The deposit also holds the quenched-and-annealed model of their Supplementary Figs.~4--5 (arXiv version), which I did not use.} with the orthogonal Kwon tight-binding Hamiltonian: 4~million orbitals, 65 nonzeros per row. At $M=300$, $R=2$ the DOS takes 58~s (plus 58~s for the spectral bounds and 18~s to build $H$) with a peak of 10.6~GB; at $M=600$, $R=4$ it takes 217~s. The band energy is $-19.882\pm0.002$ and $-19.887\pm0.003$~eV/atom for the two runs (Maxent from the saved moments agrees to $\le 0.5$~meV at every $M$ from 60 to 600, and converged in 15--16 Newton steps with the default floor, Sec.~\ref{sec:recon}); the Fermi level is $0.52$--$0.54$~eV. The DTW4096 value ($-20.038$~eV/atom, exact) belongs to a different structure and is not a reference for this model.
Fig.~\ref{fig:morrow} shows the result. The spectrum has a pseudogap, not a clean gap: between 0.1 and 0.9~eV the density stays at $(4$--$7)\times10^{-3}$ per orbital ($0.02$--$0.03$ states/atom/eV), against $\approx10^{-3}$ for DTW4096. The two runs differ by $R$ (2 vs 4) and $M$ (300 vs 600) and agree in this region to $\sim15\%$, so the level is neither stochastic noise nor Chebyshev resolution ($\pi a/M\approx0.11$~eV at $M=300$); it is a property of this structure in this Hamiltonian (with a bond cutoff of 2.85~\AA\ the deposited structure has 6\,558 threefold atoms, $0.66\%$, in line with the $0.7\%$ dangling bonds quoted by Morrow \emph{et al.}, and 16\,809 fivefold atoms, $1.68\%$, so about $2.3\%$ coordination defects in all). I did not test the Kwon Hamiltonian against a first-principles calculation of this structure.

\subsection{A million-atom replica of a 216-atom amorphous cell}
\label{sec:app_r10}

V. L. Deringer suggested a test on a periodic replica of a small defect-free amorphous model, the way Ref.~\cite{deringer} treats crystalline Si. I took the 216-atom model R10 of Rosset \emph{et al.}\ (the defect-free model of Ref.~\cite{rosset2026}, relaxed here with the same ACE potential) and tiled it $17^3$ times: 1\,061\,208 atoms, 4\,244\,832 orbitals, 3\,183\,624 vibrational degrees of freedom.
This is a stringent test for a different reason than DTW4096: the exact spectrum is known for the whole replica, since for a periodic tiling it is the union of the spectra of the cell's Bloch matrices $H(\mathbf k)$ and $D(\mathbf k)$ on the $17^3$ mesh of $\mathbf k$ points (time-reversal symmetry halves the work).
The relation is exact only while the replica stays periodic; the relaxed cell is a stable minimum (no negative eigenvalue), and a short anneal followed by minimization of a $6^3$ replica returned to the same structure, so I did not anneal the large one.
It also means that the large system contains no states other than those of the 216-atom cell at the $17^3$ $\mathbf k$ points: a larger sampling of the same network, not a larger network.
The electronic and vibrational tests below are therefore tests of the method at scale, not new physics of amorphous silicon.

\begin{figure*}
\includegraphics[width=0.85\textwidth]{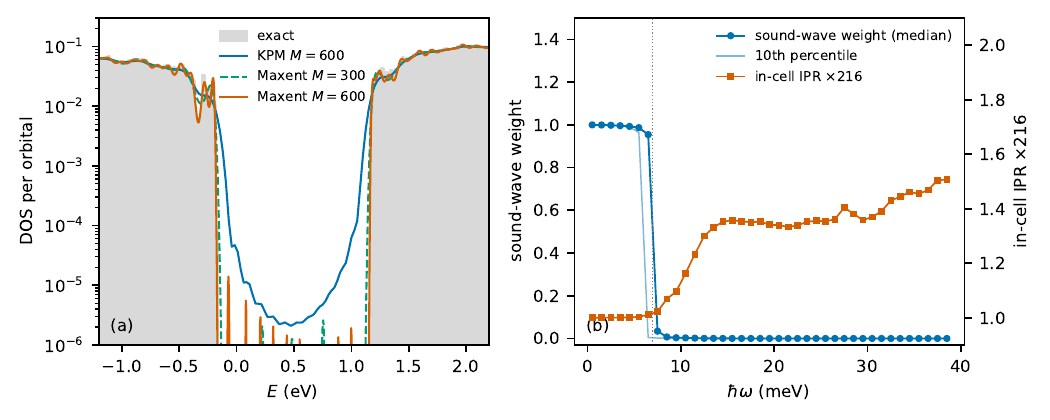}
\caption{\label{fig:r10}$17^3$ replica of the 216-atom model R10 (1\,061\,208 atoms). (a) Kwon tight-binding DOS (4\,244\,832 orbitals, $R=4$) in the gap and tails, against the exact spectrum from the Bloch matrices (0.02~eV bins). (b) Exact vibrational modes of the same replica: median and 10th percentile of the weight on a sound wave of the same wave vector, and median in-cell IPR (times 216; uniform is 1), per 1~meV bin. The dotted line (7~meV) marks the step, which lies near the zone-face frequency $\hbar v\pi/L\approx6.4$~meV of the 216-atom cell (see text); it is not a property of the network.}
\end{figure*}

\emph{Electronic structure.} With the Kwon Hamiltonian, $M=600$, $R=4$ and no exact moments the DOS takes 229~s (17~s to build $H$, peak 12.2~GB). The exact gap of the replica is $1.351$~eV ($-0.179$ to $1.172$~eV).
Figure~\ref{fig:r10}(a) shows the gap and the tails. The exact edges are sharp, because every state is one of the 216-atom cell's levels; the Maxent density with $M=300$ and 600 follows them, and drops below $10^{-6}$ per orbital within the gap except for a few narrow spikes (up to $\sim10^{-4}$ at $M=600$). KPM, at the same moment count, has a floor of $10^{-5}$--$10^{-4}$ per orbital that rises toward the band edges, so a KPM calculation at this $M$ cannot distinguish the gap from a faint pseudogap, whereas Maxent can.
In cumulative terms the largest error in $N(E)$ against the exact count is 1\,800 states out of 4\,244\,832 ($0.04\%$) for Maxent at $M=600$ and 3\,200 ($0.08\%$) for KPM; these errors sit in the band tails, not inside the bands.

\begin{figure*}
\includegraphics[width=0.85\textwidth]{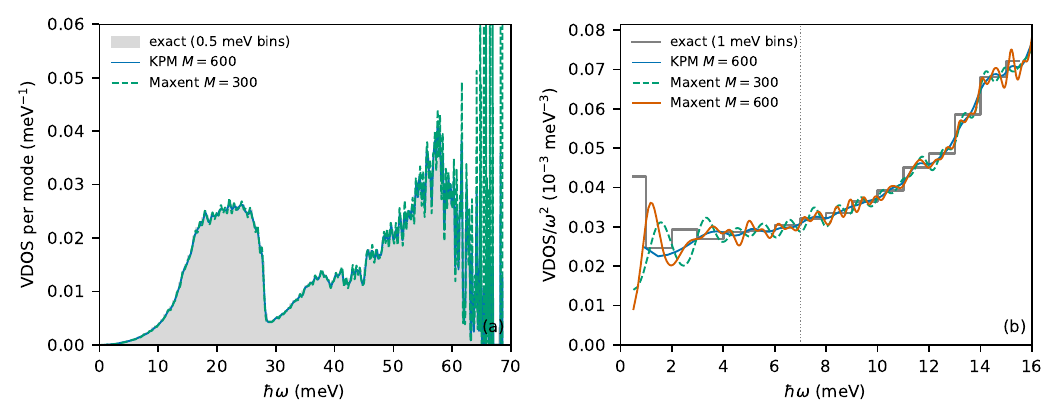}
\caption{\label{fig:r10vdos}Vibrational DOS of the $17^3$ replica of R10 from the matrix-free ACE operator ($M=600$, $R=2$) against the exact Bloch levels. (a) Whole spectrum per mode; the Maxent $M=600$ curve, which oscillates strongly above 60~meV, is omitted, and the Maxent $M=300$ curve also goes off scale above about 62~meV. (b) The low-frequency region divided by $\omega^2$ for the exact levels (gray, 1~meV bins), KPM $M=600$ (blue), Maxent $M=300$ (green, dashed) and Maxent $M=600$ (orange), as in (a); the dotted line (7~meV) is the same as in Fig.~\ref{fig:r10}.}
\end{figure*}

\emph{Vibrations.} The matrix-free ACE spectrum (\texttt{lammps} with Kokkos/CUDA on one 12~GB GPU, two force evaluations per matrix--vector product, $M=600$, $R=2$) takes 2.2~h and 1.1~GB of host memory. The zero-point energy agrees with the exact Bloch value to $0.03\%$ (KPM and Maxent, $M=170$ and 300) and $0.07\%$ (Maxent $M=600$); at temperatures above 50~K $C_v$ and $S$ agree to better than $0.1\%$. At 5~K, where only the lowest frequencies contribute, KPM at $M=600$ is off by $-18\%$ ($C_v$) and $-30\%$ ($S$), Maxent by $-3\%$ and $-2\%$.
The largest error in the integrated mode count below 9~meV, where the exact replica has 25\,297 modes, is 118 (KPM) and 130 (Maxent) at $M=600$, about $0.5\%$, but at $M=170$ it is 760 (KPM).
Figure~\ref{fig:r10vdos}(a) shows the whole VDOS. KPM and Maxent follow the exact histogram over the acoustic and optic bands, including the dip at 28~meV; above about 60~meV, where the spectrum falls to zero at 68.2~meV, both oscillate about the exact curve, and Maxent more strongly (at $M=600$ the oscillation is off the scale of the figure, and it has a spike at the exact maximum). The fraction of modes above 65~meV is nevertheless right ($1.38\%$ for Maxent, $1.41\%$ for KPM, $1.39\%$ exact). Fig.~\ref{fig:r10vdos}(b) divides by $\omega^2$: below 7~meV the exact VDOS is Debye-like, with a flat $g/\omega^2\approx3\times10^{-5}$~meV$^{-3}$ that both methods reproduce to the resolution of the fit, and above it $g/\omega^2$ rises. Below about 3~meV the reconstructions oscillate around the exact histogram, whose bins there hold only a few hundred modes.
Fig.~\ref{fig:r10}(b) shows the character of the exact modes, measured by the weight on the plane wave $e^{i\mathbf k\cdot\mathbf r}\boldsymbol\epsilon$ (summed over polarizations) with the wave vector $\mathbf k$ of the Bloch state. Below $\hbar\omega\approx6$~meV (1.5~THz) the modes are sound waves of the cell: this weight is $\ge0.99$ (median) and the in-cell participation ratio equals $1/216$, as for a uniform cell. The weight falls from $0.95$ to $0.03$ (median) between 6--7 and 7--8~meV, and above it the modes remain extended in the cell (IPR below $1.5/216$) but have no weight on the plane wave of the same $\mathbf k$.
This step is not a property of the network: it marks the Brillouin-zone boundary of the 216-atom cell. The Debye plateau of Fig.~\ref{fig:r10vdos}(b) (from the exact mode counts below 3--6~meV, $g/\omega^2=2.7$--$2.8\times10^{-5}$~meV$^{-3}$) gives $\omega_D=(3/(g/\omega^2))^{1/3}\approx48$~meV and, with 0.0477 atoms/\AA$^3$ (216 atoms in $L^3$), a mean sound velocity $v\approx5.1$~km/s ($\hbar v\approx34$~meV\,\AA), so the zone face at $\pi/L=0.19$~\AA$^{-1}$ (cubic cell, $L=16.55$~\AA) lies at $\hbar v\pi/L\approx6.4$~meV; above it the plane-wave content of a Bloch state sits at $\mathbf k+\mathbf G$, not at $\mathbf k$.
To test this I computed the same weight for DTW4096 (cell 43.7~\AA, zone face at $2.3$~meV; ten $\mathbf k$ points including $\Gamma$, 5\,049 modes below 14~meV, Bloch matrices from the Gamma-point ACE dynamical matrix, each pair assigned to its minimum image, which is valid because the force cutoff is much smaller than $L/2$; script \texttt{dtw\_bloch.py}). The weight at the same $\mathbf k$ drops from $\approx1$ to a median $0.21$ at 2--3~meV and to zero above 3~meV, i.e.\ at that cell's zone face, whereas the weight on the best plane wave $\mathbf k+\mathbf G$ (all $\mathbf G$ within two reciprocal-lattice steps, adequate below about 9~meV) decreases gradually: the medians are $0.99$, $0.97$, $0.75$, $0.69$, $0.44$, $0.25$ and $0.14$ in the 1~meV bins from 2--3 to 8--9~meV. The cell size, not the network, therefore sets the step in Fig.~\ref{fig:r10}(b), and I do not claim a propagon--diffuson crossover at 7~meV. The gradual decay of the best-plane-wave weight over roughly 4--9~meV in DTW4096 is the physical statement; I did not compute it for the replica.
\emph{Eigenvectors.} For a $3^3$ replica (17\,496 degrees of freedom) shift-invert with the Hessian assembled from finite differences found all 58 modes below 7~meV in 393~s, matching the exact eigenvalues to $10^{-8}$ (residual $3\times10^{-14}$). Shift-invert does not scale here (the ACE Hessian has about 900 nonzeros per row), so for the $17^3$ replica the modes shown in Fig.~\ref{fig:r10}(b) are Bloch eigenvectors of the cell, not shift-invert results.

\section{Cost, memory and limits}
\label{sec:cost}
\label{sec:scaling}

\paragraph*{Machine.}
All timings are from one Intel Core i5-14400F workstation (WSL2 Linux 6.18, 19~GB of memory visible, 16 logical CPUs of which at most 8 threads were used, usually 4), Python 3.14.4, NumPy 2.5.4, SciPy 1.18.1; wall-clock times of single runs.
The GPU (an RTX 5070) is used for the LAMMPS ACE dynamical matrix of DTW4096 and for the matrix-free ACE vibrational spectrum of the million-atom replica (Sec.~\ref{sec:app_r10}).
The scaling benchmarks use periodic tilings of DTW4096, \ie\ not independent structures: they measure computational cost, not physics.

\begin{figure*}
\includegraphics[width=\textwidth]{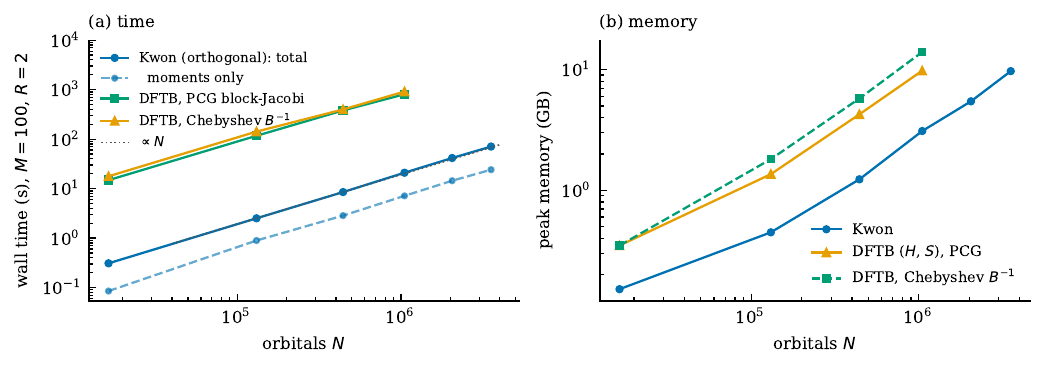}
\caption{Cost scaling on tilings of DTW4096 ($M=100$, $R=2$). (a) Wall time against the number of orbitals: orthogonal Kwon (total and moments only), non-orthogonal DFTB with PCG/block-Jacobi and with the Chebyshev $B^{-1}$; dotted: linear scaling. (b) Peak memory.}
\label{fig:scale}
\end{figure*}

\paragraph*{Orthogonal case.}
Table~\ref{tab:kw} and Fig.~\ref{fig:scale}: for the Kwon Hamiltonian on tilings up to $6^3$ copies (884\,736 atoms, 3\,538\,944 orbitals, 64 nonzeros per row) the whole DOS call ($M=100$, $R=2$) scales linearly with the number of orbitals (fitted exponent $1.013$; moments $1.05$, bounds $1.01$, matrix build $0.97$) and takes $71$~s at the largest size: $17$~s build, $29$~s bounds, $24$~s moments (100 matrix--vector products, $0.24$~s each). With the default exact moments ($K=4$) the same call takes $383$~s at the largest size ($342$~s for the moments, $14\times$ the stochastic moments of the plain run), and the moments again scale linearly with the number of orbitals (exponent $1.09$).
Peak memory is $9.7$~GB at the largest size: the CSR matrix itself is $2.7$~GB, the rest is transient copies (the scaled operator and the chunk stack), and the memory exponent $0.79$ I attribute to the fixed interpreter and library baseline at small sizes (I did not profile this); a leaner assembly would lower the peak.
The matrix-free stencil avoids storing the matrix at all (8.4 million orbitals in 1.8~GB, Sec.~\ref{sec:app_si}) but needs a translation-invariant structure.

\begin{table*}[t]\caption{Orthogonal case: Kwon tight binding on periodic tilings of DTW4096, KPM with $M=100$ moments and $R=2$ vectors, 4 threads; columns 3--7: no exact-moment substitution ($K=0$); last two columns: the default $K=4$ (moments, and the whole call including build and bounds), from a separate run. Timings are one run each.}\label{tab:kw}\begin{ruledtabular}\begin{tabular}{rrrrrrrrr}
atoms & orbitals & build & bounds & moments & total & peak mem. & moments, $K=4$ & total, $K=4$ \\ & & (s) & (s) & (s) & (s) & (GB) & (s) & (s) \\ \colrule
4\,096 & 16\,384 & 0.1 & 0.1 & 0.1 & 0.3 & 0.15 & 1.0 & 1.2 \\
32\,768 & 131\,072 & 0.6 & 1.0 & 0.9 & 2.5 & 0.45 & 8.6 & 10.0 \\
110\,592 & 442\,368 & 2.1 & 3.6 & 2.8 & 8.5 & 1.2 & 31.8 & 36.6 \\
262\,144 & 1\,048\,576 & 4.9 & 8.9 & 7.1 & 21.0 & 3.1 & 92.3 & 104.2 \\
512\,000 & 2\,048\,000 & 9.7 & 17.4 & 14.4 & 41.5 & 5.5 & 177.8 & 201.0 \\
884\,736 & 3\,538\,944 & 17.3 & 29.4 & 24.1 & 70.8 & 9.7 & 341.8 & 382.5 \\
\end{tabular}\end{ruledtabular}\end{table*}

\paragraph*{Dependence on sparseness.}
The moments cost about $\mathrm{nnz}\times M\times R/2$ multiply--adds (the doubling identities give two moments per product), so the time is linear in the number of nonzeros (Table~\ref{tab:sp}). From 15 to 255 nonzeros per row I measure $0.8$--$0.9$~ns per nonzero, moment and vector (the entry ``ns per ($\mathrm{nnz}\,MR$)'' of the table is per moment, half the time per multiply--add, i.e.\ $1.6$--$1.8$~ns per multiply--add), and up to $1.8\times$ more at 3 per row, where the per-row overhead matters.
This is a bare recursion on a random matrix without locality; the Kwon calculation above ($0.24$~s per product at $64\times3.5\times10^6$ nonzeros and $R=2$) corresponds to $\approx0.5$~ns, with the whole call also paying for the bounds and the build.
Sparseness does not change the stochastic error, which falls as $(NR)^{-1/2}$; the number of moments is set by the bandwidth divided by the resolution.
Maxent after the moments costs the same at any sparsity.
Dense exact diagonalization scales as $N^3$ and should win only for small or nearly full matrices with $MR$ comparable to $N$; I did not time that case.

\begin{table}[t]\caption{Time of the Chebyshev recursion against the number of nonzeros per row (random symmetric sparse matrix, $N=50000$, $M=100$, $R=2$ Rademacher vectors, 4 threads; bare recursion: no bounds, no reconstruction).}\label{tab:sp}\begin{ruledtabular}\begin{tabular}{rrrrrrrr}
nonzeros per row & 3 & 7 & 15 & 31 & 63 & 127 & 255 \\ \colrule
time (s) & 0.06 & 0.09 & 0.13 & 0.26 & 0.53 & 1.11 & 2.28 \\
ns per (nnz $M R$) & 1.47 & 1.07 & 0.80 & 0.80 & 0.82 & 0.87 & 0.89 \\
\end{tabular}\end{ruledtabular}\end{table}

\paragraph*{Non-orthogonal case.}
Table~\ref{tab:sk}: DFTB has 179 nonzeros per row (cutoff 6.0~\AA, 4 orbitals per atom) and a Hamiltonian and an overlap matrix, so the peak memory at the same number of orbitals is $2.3$ to $3.5$ times that of the orthogonal case with PCG and $2.3$ to $4.7$ times with the Chebyshev polynomial over the four sizes (at the largest, $3.2$ and $4.5$; ratios from the unrounded peaks in the benchmark files).
The time of the whole DOS call grows with the exponent $0.97$ (PCG, block-Jacobi, four sizes up to 1\,048\,576 orbitals) or $0.94$ (Chebyshev polynomial, four sizes) and the CG iteration count per solve is independent of size ($23.7$ at 16\,384 orbitals, $23.9$ at 442\,368 and at 1\,048\,576), as it should be for a fixed local preconditioner of a well-conditioned overlap.
IC(0) reduces the iterations to 11 per solve, but my incomplete Cholesky is a Python loop per column and takes longer overall, so I ran it only for the small sizes. A compiled factorization would change that conclusion.
A million-atom non-orthogonal DFTB problem does \emph{not} fit in 19~GB as explicit $H$ and $S$: at 262\,144 atoms the PCG run peaked at $9.8$~GB and the Chebyshev run at $14.0$~GB (912~s, under a 16~GB cap; the first attempt, under a 10~GB cap, was killed), so four times that size would need $40$--$56$~GB.

\begin{table*}[t]\caption{Non-orthogonal case: DFTB ($pbc$-0-3, non-self-consistent) $H$ and $S$ on tilings of DTW4096, KPM with $M=100$, $R=2$; time of the whole density-of-states call for the three ways of applying $S^{-1}$ (PCG with block-Jacobi, Chebyshev polynomial, PCG with IC(0)). IC(0) was run only for the small sizes (its setup is a Python loop). At 262\,144 atoms the Chebyshev row is a rerun under a 16~GB cap (peak 14.0~GB; the first attempt, under a 10~GB cap, was killed); the PCG row at that size was recovered from the log of the first run (peak 9.8~GB), and the build time is that of the rerun.}\label{tab:sk}\begin{ruledtabular}\begin{tabular}{rrrrrrrrr}
atoms & orbitals & nnz/row & build & PCG (BJ) & Chebyshev & PCG (IC(0)) & peak mem., PCG & peak mem., Cheb. \\ & & & (s) & (s) & (s) & (s) & (GB) & (GB) \\ \colrule
4\,096 & 16\,384 & 179 & 1 & 15 & 18 & 35 & 0.35 & 0.35 \\
32\,768 & 131\,072 & 179 & 4 & 117 & 142 & 288 & 1.4 & 1.8 \\
110\,592 & 442\,368 & 179 & 15 & 380 & 398 & -- & 4.3 & 5.8 \\
262\,144 & 1\,048\,576 & 179 & 26 & 806 & 912 & -- & 9.8 & 14.0 \\
\end{tabular}\end{ruledtabular}\end{table*}

\paragraph*{Vibrations.}
A matrix-free product costs two force evaluations. With Stillinger--Weber on one CPU thread a $12\,288$-degree-of-freedom model takes $5$~ms per force call (11~s for $M=100$, $R=2$ and the bounds) and $98\,304$ degrees of freedom $37$~ms per call ($191$~s, 5124 calls including the bounds), \ie\ linear in the force cost. Only about 200 of these calls belong to the moments ($M=100$, $R=2$); about 4900 (some 2450 Lanczos products) go to the spectral bounds and the Hermiticity probes. In the 4096-atom run the bounds took $9.0$~s of $9.9$~s and the 102 products of the moments $0.8$~s, so the bounds dominate the cost of a matrix-free vibrational density of states at these sizes. I did not investigate why ARPACK needs so many products here; it is not the finite-difference step (changing it from $10^{-3}$ to $10^{-5}$~\AA\ changes the count of force calls for the bounds of the 4096-atom model only from 2124 to 2584), and the window found is correct (Stillinger--Weber, $[0,1.844]$).
With ACE the force call dominates: on CPU it takes $0.73$~s for 4096 atoms (4 threads) and the matrix-free DOS of DTW4096 took 1306~s for 1768 force calls; with a Kokkos/CUDA build of LAMMPS on one 12~GB GPU through its Python module the million-atom replica of Sec.~\ref{sec:app_r10} (1\,220 force calls for $M=600$, $R=2$, with the spectral window taken from the exact force-constant spectrum) took 2.2~h, i.e.\ about 6.6~s per call including the transfer of positions and forces through Python, and 1.1~GB of host memory.

\paragraph*{What does not scale.}
(i) \emph{Shift-invert and the exact HOMO--LUMO bisection} are not $O(N)$ because of fill; for the Xu fullerenes the bisection takes 27~s at 15\,360 orbitals and $62$--$80$~s at 19\,440--24\,000, and the exact-gap bisection costs more than the DOS above $\sim2\times10^4$ orbitals.
(ii) \emph{Exact moments} of order $K$ cost a time that grows quickly with $K$ (1.8, 9 and 80~s per run at $K=4$, 8, 16 for 16\,384 orbitals), although it is linear in $N$ for fixed $K$.
(iii) The \emph{spectral bounds} are not negligible and grow with $N$ (162 to 522 matrix--vector products from 16\,384 to $3.5\times10^6$ orbitals of the stencil, against 340 for the moments with $M=170$, $R=4$); for a crystal they can be given analytically. For a matrix-free operator they can also be wrong: on a relaxed one-million-atom ACE Hessian (the Morrow \emph{et al.}\ model of Sec.~\ref{sec:app_dtw}, whose vibrational density of states is not part of this report; finite-difference step $10^{-3}$~\AA) the Lanczos window came out as $[-29,69]$ in units of eV\,\AA$^{-2}$\,amu$^{-1}$, whereas the spectrum lies in about $[0,1.34]$ (a plain 30-step Lanczos iteration with full reorthogonalization and a step of $10^{-4}$~\AA\ converges to extreme Ritz values $0.0035$ and $1.336$; the exact ranges of the 216-atom cell replica and of DTW4096 are $[0,1.11]$ and $[0,1.27]$). The cause is the finite-difference step: the displacement is normalized to an rms of $10^{-3}$~\AA\ over all $3N$ components, so a Lanczos vector that localizes on a few atoms is displaced by a large amount per atom, the product is then not that of a fixed linear operator, and the Ritz values leave the spectrum (the same iteration with the step of $10^{-3}$~\AA\ reaches $[-5.3,20]$ after 20 steps). I inferred the localization from this step dependence and did not measure it. With a window of $[-29,69]$ and $M=600$ moments the resolution is $0.26$, a quarter of the band, and the result is useless; the symptom is a density with weight at $\lambda<0$ (the result is flagged \texttt{negative\_indicated}) and a window much wider than the physical range; a window that is too narrow is as bad (the moments of a run with $[-0.05,1.3]$ grew to $10^{179}$). For matrix-free vibrational problems I therefore recommend estimating the window with a smaller step (here $10^{-4}$~\AA) or passing \texttt{bounds} from a known range (a crystal, or a small cell of the same potential), and checking the window before a long run; the runs of this report that used it (Sec.~\ref{sec:app_r10}) took the window from the exact spectrum of the force-constant matrix.
(iv) \emph{Maxent with many exact moments} stops converging unless the moments are given a small uncertainty (Sec.~\ref{sec:val_acc}). Even then its pointwise $L_1$ error against the 0.05~eV-broadened exact spectrum is smallest at $M=170$ and rises at $M=300$ and 400, while KPM's keeps falling (Table~\ref{tab:acc}). The floor is not the cause (Sec.~\ref{sec:val_acc}); part of it is the test (a sharp Maxent curve is compared with a broadened spectrum), and for the band energy, an integrated quantity, Maxent is the more accurate at every $M$ in Table~\ref{tab:acc}. I do not claim that a converged Maxent is worse than KPM in general: for integrated quantities it is better, and only for pointwise resolution at large $M$ do I recommend KPM.
(v) The \emph{DFTB matrices} are large: 179 nonzeros per row.
(vi) The statistical error of a moment decreases with $\sqrt{NR}$, so large systems need few vectors, but the \emph{band energy per atom of one 4096-atom cell with $R=6$ vectors has an uncertainty of order 0.1~eV/atom} (Table~\ref{tab:dd}; rms 110~meV/atom at $R=6$) unless exact moments or impartial vectors are used (Table~\ref{tab:vec}), and for the generalized problem the variance reductions of Table~\ref{tab:vec} are not available; integrated quantities should be reported with their error.
(vii) Not checked against an independent reference: the complex generalized problem at a $\mathbf k$ point (Sec.~\ref{sec:val_cplx}), the generalized problem through the result file (Sec.~\ref{sec:impl}), the Mulliken weights and IPR (tested through sum rules and one shift-invert case), and the one-million-atom amorphous run.
(viii) In the replica of Sec.~\ref{sec:app_r10} Maxent oscillates strongly where the vibrational spectrum ends abruptly (above 60~meV at $M=600$), although the integrated weight is right; shift-invert is practical only for small replicas, and the exact references there exist because the replica is periodic.
(ix) Finally, all numbers are $\Gamma$-point or single-twist values for one structure each, except the crystal and replica comparisons, whose references are exact $\mathbf k$-sums; the dependence on the structure and the cell size is a separate question.

\section{Recommendations}
\label{sec:recs}

\begin{enumerate}
\item \textbf{KPM is the default}: linear, positive (Jackson), never fails, resolution $\pi a/M$; for the maximum resolution of a large system increase $M$ (a few thousand is cheap; cost is linear in $M$).
Beyond the mean level spacing or the noise floor, more moments only sharpen sampling speckle.
\item \textbf{Maxent} (keep the default uncertainty floor, which makes it converge with exact or nearly exact moments) when high-order moments are expensive or noisy ($M\lesssim300$), when positivity and sharp edges matter, and for integrated quantities (band energy, gap region); from identical exact moments at $M=20$--70 (the sizes where it converged) it gave 8--12$\times$ lower band-energy errors and 6--30$\times$ lower $L_1$ distances (Table~\ref{tab:acc}).
Use the bootstrapped solver; with the default uncertainty floor it converges also on exact moments, but its pointwise error, measured against a broadened exact spectrum, rises beyond a few hundred moments of singular densities (Table~\ref{tab:acc}).
\item \textbf{Random vectors}: the default exact-moment substitution ($K=4$) is almost always worth it for KPM, and for Maxent with $R\ge2$ vectors; \emph{with Maxent and a single vector use $K=0$ or impartial vectors} (Sec.~\ref{sec:small}); $K=8$--16 for band energies; impartial vectors constrain only three moments and are nearly as efficient at lower cost per vector (Table~\ref{tab:vec}).
\item \textbf{Generalized problems}: PCG with block-Jacobi is the default and my plain recommendation; the Chebyshev polynomial needs no inner solves, which makes it the option for matrix-free or GPU settings where no iterative-solver stack is wanted, but in Table~\ref{tab:sk} it was slower and needed more memory than PCG; IC(0) lowers the iteration count only if its setup is cheap (mine is not).
\item Report the \textbf{sampling error} of every integrated quantity, and overlay KPM and Maxent: where they agree, the feature is in the moments.
\end{enumerate}

\section{Conclusions and availability}
\label{sec:conc}

\sdos\ computes densities of states and related spectral quantities of very large sparse Hermitian and generalized problems.
The ingredients are established: stochastic traces, KPM, Maxent, and non-orthogonal KPM by PCG. What is added is their combination, the choices for the overlap inverse, bootstrapped Maxent, windows, projections, the interfaces, and a validation against exact diagonalization and independent codes from $10^3$ to $8\times10^6$ orbitals, including the exact Bloch spectra of a million-atom replica of a 216-atom amorphous cell, for which Maxent resolves the gap edges that KPM smears and the matrix-free ACE vibrational spectrum reproduces the zero-point energy to 0.07\%.
The moment-based computations scale linearly in the number of orbitals for explicit matrices (exponent $1.01$) and nearly so for the matrix-free crystal ($1.12$). Three conventions of the DFTB file format that I did not find in the documentation available to me are recorded here. The limits are stated in Sec.~\ref{sec:scaling}.

\paragraph*{Availability.}
The package is licensed under BSD-3-Clause (copyright D.~A.~Drabold, 2026) and is available at \url{https://github.com/Drabold/sdos} and archived on Zenodo (version 0.1.0, concept DOI \href{https://doi.org/10.5281/zenodo.24023874}{10.5281/zenodo.24023874}); it installs with \texttt{pip install sdos} (PyPI) or from a clone with \texttt{pip install -e .}. The repository contains the examples and benchmark scripts of this report; the large run data (million-atom structures and exact spectra) and the DFTB parameter files are not included.
All scripts, saved results and the generation of the figures and tables of this report are in the \texttt{examples/} and \texttt{paper/} directories of the repository; the tests run with \texttt{pytest}.

\begin{acknowledgments}
AI assistance: the package code, the tests and the benchmark and analysis scripts were written, and the text of this report was drafted, with Claude Sonnet 5.5 and Claude Opus 5.5 (Anthropic), used through Claude Code. I reviewed and validated all AI-assisted output and made the design decisions; I am responsible for all scientific statements.
I acknowledge the Frauenheim group's \texttt{pbc-0-3} Slater--Koster set (CC BY-SA~4.0), whose required references I cite.
I thank R. N. Silver, H. R\"oder, J. J. Dong, P. Ordej\'on and R. M. Martin for their insights and collaboration, C. Ugwumadu for a careful reading of the whole manuscript and many helpful suggestions, E. T. Drabold for advice on hardware selection, V. L. Deringer for suggesting the million-atom replica test, and L. A. M. Rosset and V. L. Deringer for the silicon--hydrogen ACE potential used in the benchmarks.
\end{acknowledgments}

\bibliography{sdos}

\end{document}